\documentclass[a4paper,fleqn,usenatbib]{mnras}

\usepackage{txfonts}

\usepackage[T1]{fontenc}
\usepackage{ae,aecompl}

\usepackage{psfig}   
\usepackage{graphicx}
\usepackage{amssymb}
\usepackage{subfigure}

\title[Planets in binaries]{Long-term stability of planets in and around binary stars\thanks{A video abstract for this paper created by Helena Gibbon is available at: \href{https://youtu.be/76kANnTK9-s}{https://youtu.be/76kANnTK9-s}}}% systems}

\author[H. A. Ballantyne, T. Espaas et al.]{Harry A. Ballantyne$^{1,2}$, %\thanks{harry.ballantyne@space.unibe.ch}
 Tore Espaas$^{1}$, Bethan Z. Norgrove$^1$, Bethany A. Wootton$^1$, 
\newauthor \hspace*{-0.15cm} Benjamin R. Harris$^1$, Isaac L. Pepper$^1$, Richard D. Smith$^3$, Rosie E. Dommett$^1$ and \newauthor \hspace*{-0.15cm} Richard  J. Parker$^1$\thanks{E-mail: R.Parker@sheffield.ac.uk}\thanks{Royal Society Dorothy Hodgkin Fellow} \vspace*{0.1cm}\\
$^1$Department of Physics and Astronomy, The University of Sheffield, Hicks Building, Hounsfield Road, Sheffield, S3 7RH, UK\\
$^2$Universit{\"a}t Bern, Space Research \& Planetary Sciences (WP), Gesellschaftsstrasse 6, 3012 Bern, Switzerland\\
$^3$Astrophysics Research Centre, School of Mathematics and Physics, Queen's University Belfast, Belfast, BT7 1NN, UK}

\begin{document}

\date{}
                             
\pagerange{\pageref{firstpage}--\pageref{lastpage}} \pubyear{2018}

\maketitle

\label{firstpage}

\begin{abstract}
Planets are observed to orbit the component star(s) of stellar binary systems on so-called circumprimary or circumsecondary orbits, as well as around the entire binary system on so-called circumbinary orbits. Depending on the orbital parameters of the binary system a planet will be dynamically stable if it orbits within some critical separation of the semimajor axis in the circumprimary case, or beyond some critical separation for the circumbinary case. We present $N$-body simulations of star-forming regions that contain populations of primordial binaries to determine the fraction of binary systems that can host stable planets at various semimajor axes, and how this fraction of stable systems evolves over time. Dynamical encounters in star-forming regions can alter the orbits of some binary systems, which can induce long-term dynamical instabilities in the planetary system and can even change the size of the habitable zone(s) of the component stars. However, the overall fraction of binaries that can host stable planetary systems is not greatly affected by either the assumed binary population, or the density of the star-forming region. Instead, the critical factor in determining how many stable planetary systems exist in the Galaxy is the stellar binary fraction -- the more stars that are born as singles in stellar nurseries, the higher the fraction of stable planetary systems. 
\end{abstract}


%\section{Outline}

%\begin{itemize}

%\item majority of star formation results in binary or multiple systems

%\item Planets form concurrently with stars, and we observe both S-type and P-type planetary systems.

%\item Many studies have demonstrated that phase space in binary orbits can be stable for planetary systems to form and evolve.

%\item However, many studies have shown that the dense environments in which (binary) stars form  can be disruptive for planet formation. Direct encounters with passing stars can liberate planets from single stars. So-called secondary effects, where a binary star (with planets) experiences an encounter, can lead to long-term destabilising of the planetary systems. E.g. Von Zeipel-Lidov-Kozai. A previously stable planet can be made unstable following an interaction, and a previously unstable planet could be made stable (or in reality, the phase space in which a planet could be stable increases).

%\end {itemize}

\section{Introduction}

The majority of star formation results in binary or higher-order multiple systems \citep{Duquennoy91,Connelley08,Raghavan10,Chen13,Pineda15,Duchene13a}, with some studies suggesting that the initial multiplicity fraction is close to unity. 

Furthermore, most stellar systems (both single and multiple) form in groups where the stellar density can exceed that in the Galactic field by  factors of 10 -- 10\,000 \citep{Lada03,Gutermuth09,Bressert10}. These groups may form short-lived unbound stellar associations \citep{Blaauw61} or loose moving groups \citep{Zuckerman04,Gagne18b}, or may form longer-lived gravitationally bound clusters \citep{Kruijssen12b}. The relatively high stellar densities and longevity of bound clusters means that dynamical encounters are likely to be important in these environments, although even an unbound association can in principle be dense at the time of its formation \citep{Parker14b}.

In the densest environments, dynamical encounters between passing stars are common, and can significantly alter the orbits of binary and multiple systems \citep{Kroupa95a,Kroupa95b,Kroupa95c,Kroupa99,Fregeau04,Fregeau06,Parker11c,Marks12,Cloutier21}. The overall binary fraction is reduced by these encounters \citep[despite a small proportion of new binaries forming due to the dissolution of star-forming regions,][]{Kouwenhoven10,Moeckel10}, and orbital parameters such as the semimajor axis, eccentricity and inclination can be drastically altered.  

However, at the same time as (single and multiple) stars form, planets are observed to be forming concurrently. This is inferred from both the presence of protoplanetary discs around young stars \citep[e.g.][]{Haisch01,Richert18}, as well as direct signatures of planet formation in discs that have been imaged by ALMA \citep{Brogan15,Andrews18,Alves20,SeguraCox20}. Planets are observed to form in and around binary stars; either orbiting the primary and/or secondary components on a satellite, or S-type orbit, or orbiting the entire binary on a so-called planetary, or P-type orbit \citep{Dvorak86}. 

If planets are able to form in and around binary systems, this implies that the binary system must be dynamically stable for long periods of time. \citet{Holman99} derived analytic formulae from the results of $N$-body simulations of mass-less particles in binary systems, and these relations provide an estimate of the stability of a binary system, given the component masses, semimajor axis and eccentricity of the binary. 

In addition, several authors have explored the effects of `secondary' dynamic encounters; encounters that do not initially destroy the planetary system within a binary, but lead to conditions that later destabilise the planet(s). One such mechanism is the von~Zeipel-Lidov-Kozai mechanism \citep{vonZeipel10,Lidov62,Kozai62}, which is a transfer of angular momentum onto planets within a binary system, which cause oscillations in the inclination and eccentricity of the binary \citep{Innanen97,Fabrycky07}. Several studies have demonstrated that the vZLK mechanism can lead to planet-planet scattering and ejection \citep[e.g.][]{Malmberg07a}, and the inward migration of Jupiter-mass planets \citep{Wu03}.  

Given that dynamical encounters can alter the orbits of binary systems, and that planets form early on in the star formation process, a significant fraction of binary systems that host stable planets could be made unstable by encounters in dense star-forming regions. In previous work \citep{Wootton19} we have demonstrated that dynamical encounters can push close ($<10$\,au) binaries together, so that  the size of the habitable zone around the binary system increases. In this paper, we examine the effects of dynamical encounters on the long-term stability of planets in and around binary systems to determine the change (if significant) in the fraction of binary systems that could host stable planets.

The paper is organised as follows. In Section~\ref{method} we outline the method for setting up $N$-body simulations of the evolution of star-forming regions with various primordial binary populations, as well as the method for determining the stability of planets and the calculation to determine the size of the habitable zone. In Section~\ref{results} we present our results, we provide a discussion in Section~\ref{discuss} and we conclude in Section~\ref{conclude}. 

%\begin{figure*}
 % \begin{center}
%\setlength{\subfigcapskip}{10pt}
%%\hspace*{0.3cm}\subfigure[$D = 1.6$]{\label{}\rotatebox{270}{\includegraphics[scale=0.35]{plot_fbin_Or_C0p3F1p61pB_F10.ps}}}
%%\hspace*{0.5cm}\subfigure[$D = 3.0$]{\label{}\rotatebox{270}{\includegraphics[scale=0.35]{plot_fbin_Or_C0p3F3p01pB_F10.ps}}}
%\hspace*{0.3cm}\subfigure[$D = 1.6$]{\label{}\rotatebox{270}{\includegraphics[scale=0.35]{plot_fbin_Or_C0p3F1p61pRmR10.ps}}}
%\hspace*{0.5cm}\subfigure[$D = 3.0$]{\label{}\rotatebox{270}{\includegraphics[scale=0.35]{plot_fbin_Or_C0p3F3p01pRmR10.ps}}}
%\caption[bf]{Evolution of binary fractions. } 
%\label{}
%  \end{center}
%\end{figure*}

%\section{Habitable zones in binary stars}

%\section{Dynamical evolution of binary orbits in star-forming regions}

\section{Method}
\label{method}

In this section we describe the $N$-body simulations used to model the star-forming regions\footnote{We use the terminology `star-forming region' because at $t = 0$\,Myr these simulations are not yet star clusters in the classical sense. However, they are devoid of gas at all times.}, the set-up of the binary systems and the criteria used to determine the semimajor axis range within which a planet(s) could exist on stable orbits  in and around our binary systems.

\subsection{Star-forming regions}

The distributions of  pre-main sequence stars in star-forming regions are generally observed to exhibit spatial and kinematic substructure \citep{Gomez93,Larson95,Cartwright04,Hacar17}. Furthermore, the overall (bulk) motion of objects in star-forming regions is observed to be subvirial (collapsing) for pre- and protostellar cores, but becoming virialised for older objects like pre-main sequence stars \citep{Foster15}. This has led to the interpretation that star-forming regions are initially subvirial, and then collapse to form smooth, centrally concentrated clusters \citep{Allison10,Parker14b,Parker16b}. This would also be apparent in the spatial distributions of stars, which would attain smoother morphologies after several dynamical timescales \citep[e.g.][]{DaffernPowell20} as the substructure is erased, and in the velocity dispersions, which would appear more virialised at later times \citep{Parker16b}.

Following this, we set our star-forming regions up to be subvirial, where the virial ratio is defined as
\begin{equation}
\alpha_{\rm vir} = T/|\Omega|,
\end{equation}
where $T$ and $|\Omega|$ are the total kinetic and potential energies of the stars. A region in virial equilibrium has $\alpha_{\rm vir} = 0.5$; our subvirial regions have $\alpha_{\rm vir} = 0.3$.  This choice of virial ratio is somewhat arbitrary; we wish to mimic the subvirial nature of star formation but lower initial virial ratios (i.e. even more subvirial) would increase the amount of violent relaxation and could lead to more binary processing. However, we note that most of the binary processing occurs within the substructure, and so the effect of the virial ratio is of secondary importance to the local density, as set by the substructure \citep{Parker11c}.

To mimic the primordial substructure in star-forming regions we use the box-fractal distribution described in \citet{Goodwin04a}. These models have the advantage that the amount of spatial and kinematic substructure is set by just one number, the fractal dimension $D$. However, these models are unlikely to capture all of the aspects of spatial and kinematic structure in young star-forming regions, and we refer the interested reader to \citet{Sills18} for alternative initial conditions for star-forming regions. 

The construction of the fractals is outlined in our previous work \citep[e.g][]{Allison10,Parker14b,DaffernPowell20} and so we only briefly describe them here. The fractal dimension quantifies the amount of substructure; regions with $D = 1.6$ have a high degree of substructure, and regions with $D = 3.0$ are smooth. The box-fractal method places a `parent' particle at the centre of a cube, and the probability of that particle maturing and producing further `child' particles is given by the fractal dimension. The higher the fractal dimension, the more particles mature and so there is less substructure. 

Velocities of the parent particles are drawn from a Gaussian distribution with a mean of zero, and the children inherit these velocities plus a small random offset that decreases in magnitude for successive generations of particles. Finally, the velocities are then scaled to the bulk virial ratio ($\alpha_{\rm vir} = 0.3$).

We fix the initial radii of our fractals to be $r_F = 1$\,pc. For the same number of stars (we use $N = 1500$, see next subsection), a region will have a very different \emph{average local density} depending on the fractal dimension. For example, our substructured regions with $D = 1.6$ have a median local stellar density of $\tilde{\rho} \sim 10\,000$\,M$_\odot$\,pc$^{-3}$, because the stars are exclusively located in clumps of substructure, whereas our smooth regions  with $D = 3.0$ have a median local stellar density of $\tilde{\rho} \sim 100$\,M$_\odot$\,pc$^{-3}$, because the stars are uniformly distributed.

\subsection{Binary populations}

Our star-forming regions contain $N = 1500$ stars in total. We draw primary masses from the \citet{Maschberger13} Initial Mass Function, which has a probability density function of the form:
\begin{equation}
p(m) \propto \left(\frac{m}{\mu}\right)^{-\alpha}\left(1 + \left(\frac{m}{\mu}\right)^{1 - \alpha}\right)^{-\beta},
\label{maschberger_imf}
\end{equation}
where $\mu = 0.2$\,M$_\odot$ is the average stellar mass, $\alpha = 2.3$ is the \citet{Salpeter55} power-law exponent for higher mass stars, and $\beta = 1.4$ describes the slope of the IMF for low-mass objects \citep*[which also deviates from the log-normal form;][]{Bastian10}. We randomly sample this distribution in the mass range 0.02 -- 50\,M$_\odot$.

For each binary system the primary star mass $m_p$ is drawn from the above IMF, and the secondary mass $m_s$ is drawn from a flat distribution so that all values of the binary mass ratio
\begin{equation}
q = \frac{m_s}{m_p}
\end{equation}
between 0 and 1 are equally probable, in accordance with the observations of binaries in the Galactic field \citep{Reggiani11a,Reggiani13}. More recently, \citet{Moe17} argue that the companion mass ratio distribution of Solar-type binaries in the field is slightly skewed to mass ratios of unity, especially for shorter periods (smaller separations), but we adopt a flat mass ratio distribution in our simulations. Dynamical evolution does not drastically alter the mass ratio distribution \citep{Fregeau04,Fregeau06,Parker13b} and so the choice of distribution is unimportant.

We also draw binary eccentricities from a flat distribution, again in accordance with observations of binary stars in the field, save for binaries with separations less than 1\,au, which are observed to be nearly circular \citep{Duquennoy91,Raghavan10}, and are thought to tidally circularise on short timescales \citep{Zahn77,Zahn89b}. For binaries with separations less than 1\,au, we set the eccentricity to be zero.

We run simulations with two very different initial binary populations in terms of the overall binary fraction and the initial semimajor axis distribution. In our `default' simulations, the stars have a binary fraction in accordance with observations of stars in the field, where the binary fraction is defined as 
\begin{equation}
f_{\rm bin} = \frac{B}{S + B},  
\end{equation}
where $S$ and $B$ are the respective numbers of single and binary systems. We do not include triples or higher-order multiple systems in our simulations, although such systems are observed to be common in the Galactic field \citep[e.g.][]{Tokovinin08,Tokovinin14,Tokovinin18} and are probably ubiquitous in star formation \citep{Reipurth14,Pineda15}. However, for the purposes of this paper we are only interested in planets in and around binary systems, and so we neglect these more complicated multiple systems. In the Galactic field, the binary fraction appears to be a strong function of the primary mass of the system \citep[though see][]{DeRosa14} and we implement a binary fraction $f_{\rm bin} = 1.00$ for massive stars \citep[$m_p>3.0$\,M$_\odot$,][]{Sana13}, $f_{\rm bin} = 0.48$ for A-type stars \citep[$1.5 < m_p/{\rm M_\odot} \leq 3.0$,][]{DeRosa14}, $f_{\rm bin} = 0.46$ for G-, F- and K-stars  \citep[$0.45 < m_p/{\rm M_\odot} \leq 1.5$,][]{Raghavan10}, $f_{\rm bin} = 0.34$ for M-stars  \citep[$0.08 < m_p/{\rm M_\odot} \leq 0.45$,][]{Bergfors10,Janson12} and a binary fraction of $f_{\rm bin} = 0.15$ for brown dwarfs  \citep[$0.02 < m_p/{\rm M_\odot} \leq 0.08$,][]{Burgasser07}.

For our simulations with a Field-like binary population, the input semimajor axis distribution is also a function of the primary mass of each binary. Binaries with primary stars more massive than $m_p>3.0$\,M$_\odot$ have separations drawn from a log-uniform, or \citet{Opik24} distribution, which is uniform in log-space between 0 and 50\,au. For all other binaries we draw semimajor axes from a log-normal distribution, whose mean (i.e. peak semimajor axis) and variance depends on the primary mass star in the binary. For A-type stars we set the mean to be $\bar{a} = 389$\,au and variance $\sigma_{\rm log\,\bar{a}} = 0.79$ \citep{DeRosa14}, for F-, G- and K-type stars we set the mean to be $\bar{a} = 50$\,au and variance $\sigma_{\rm log\,\bar{a}} = 1.68$ \citep{Raghavan10}, for M-type stars we set the mean to be $\bar{a} = 16$\,au and variance $\sigma_{\rm log\,\bar{a}} = 0.80$ (\citealp[][]{Bergfors10,Janson12}; see also \citealp{Ward-Duong15}) and for brown dwarfs we set the mean to be  $\bar{a} = 4.6$\,au and variance $\sigma_{\rm log\,\bar{a}} = 0.4$ \citep{Burgasser07,Thies07}. We summarise the properties of the Galactic field population of binaries (which we use as inputs for the simulations) in Table~\ref{field_props}.

We also run a set of simulations with a `universal' initial binary population \citep{Kroupa95a,Kroupa95b,Kroupa08}, in which the initial binary population (in terms of the binary fraction, semimajor axis distribution, etc.) is assumed to be the same in all star-forming regions, and any differences between regions are postulated to be due to different amounts of dynamical destruction in star-forming regions of differing density \citep{Marks14}. This universal binary population was shown to be inconsistent with the observed present-day density and levels of spatial substructure in several star-forming regions \citep{Parker14e}, because the presence of substructure places strong constraints on the amount of dynamical processing that can have occurred \citep{Scally02,Goodwin04a,Parker12d,Parker14b}. However, to explore a larger parameter space, we include a set of simulations with these initial binary properties. 

For the `universal' binary populations, all stars are initially placed in binary systems $f_{\rm bin} = 1.00$, and semimajor axes are drawn from the following distribution, which is an adaptation of the initial period distribution defined in \citet{Kroupa95a}:
\begin{equation}
f\left({\rm log_{10}}a\right) = \eta\frac{{\rm log_{10}} a - {\rm log_{10}} a_{\rm min}}{\delta + \left({\rm log_{10}} a - {\rm log_{10}} a_{\rm min}\right)^2},
\label{coma}
\end{equation}
where ${\rm log_{10}} a$ is the logarithm of the semi-major axis in au and ${\rm log_{10}} a_{\rm min} = -2$ ($a_{\rm min} = 0.01$\,au). The numerical constants are $\eta = 5.25$ and $\delta = 77$. By design, this distribution contains a large number of wide $>1000$\,au binaries. In the type of dense star-forming regions we model here ($\tilde{\rho} \sim 100 - 10\,000$\,M$_\odot$\,pc$^{-3}$) many of these binaries will not be gravitationally bound, even before dynamical evolution has occurred, due to the close proximity of random members of the star-forming region. As such, the actual initial binary fraction can be much lower than unity \citep[it is around $f_{\rm bin} \sim 0.7$ for the very substructured ($D = 1.6$) regions,][]{Parker14d}.

\begin{figure*}
  \begin{center}
\setlength{\subfigcapskip}{10pt}
\hspace*{0.3cm}\subfigure[]{\label{binary_props-a}\rotatebox{270}{\includegraphics[scale=0.35]{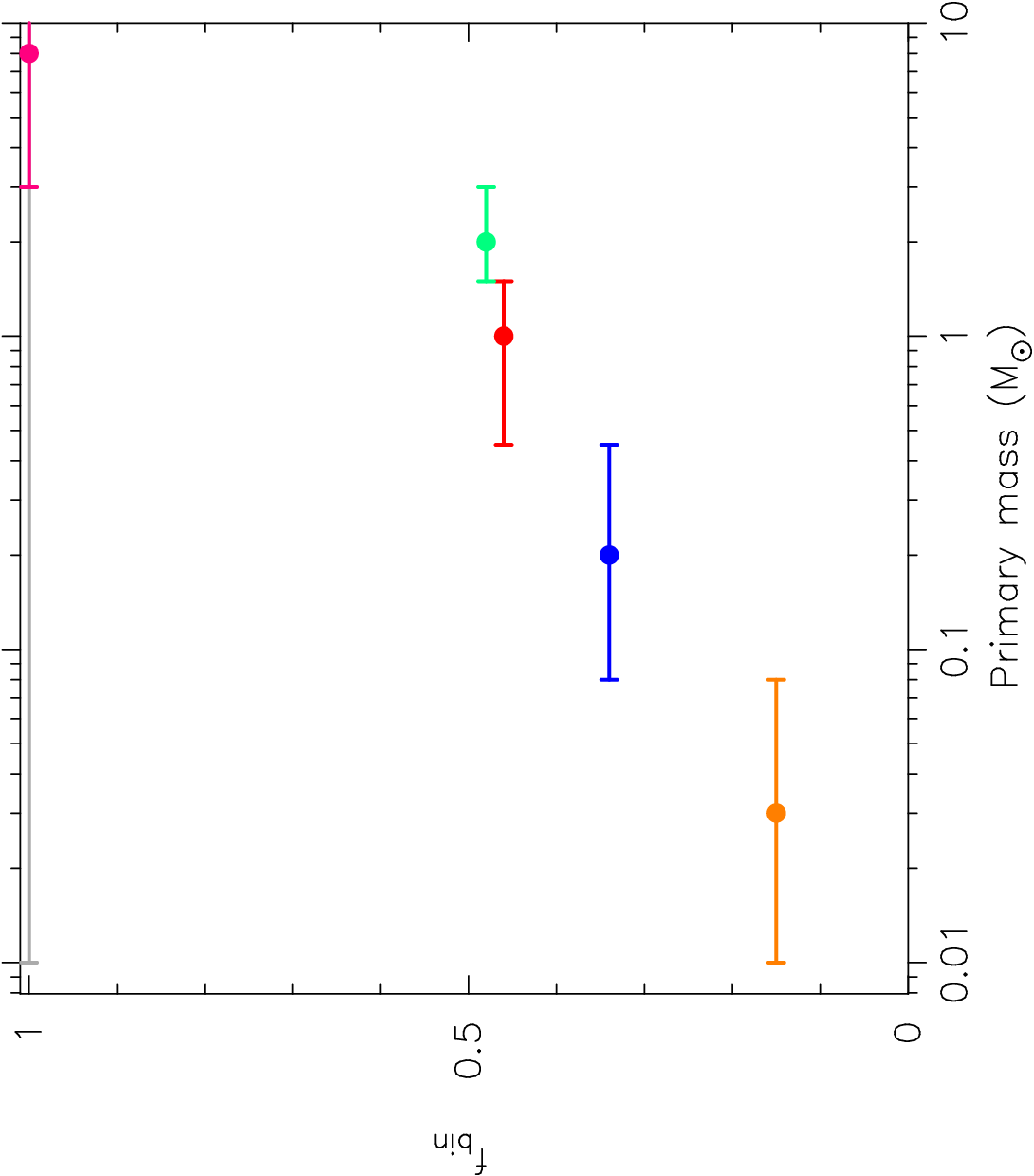}}}
\hspace*{0.5cm}\subfigure[]{\label{binary_props-b}\rotatebox{270}{\includegraphics[scale=0.35]{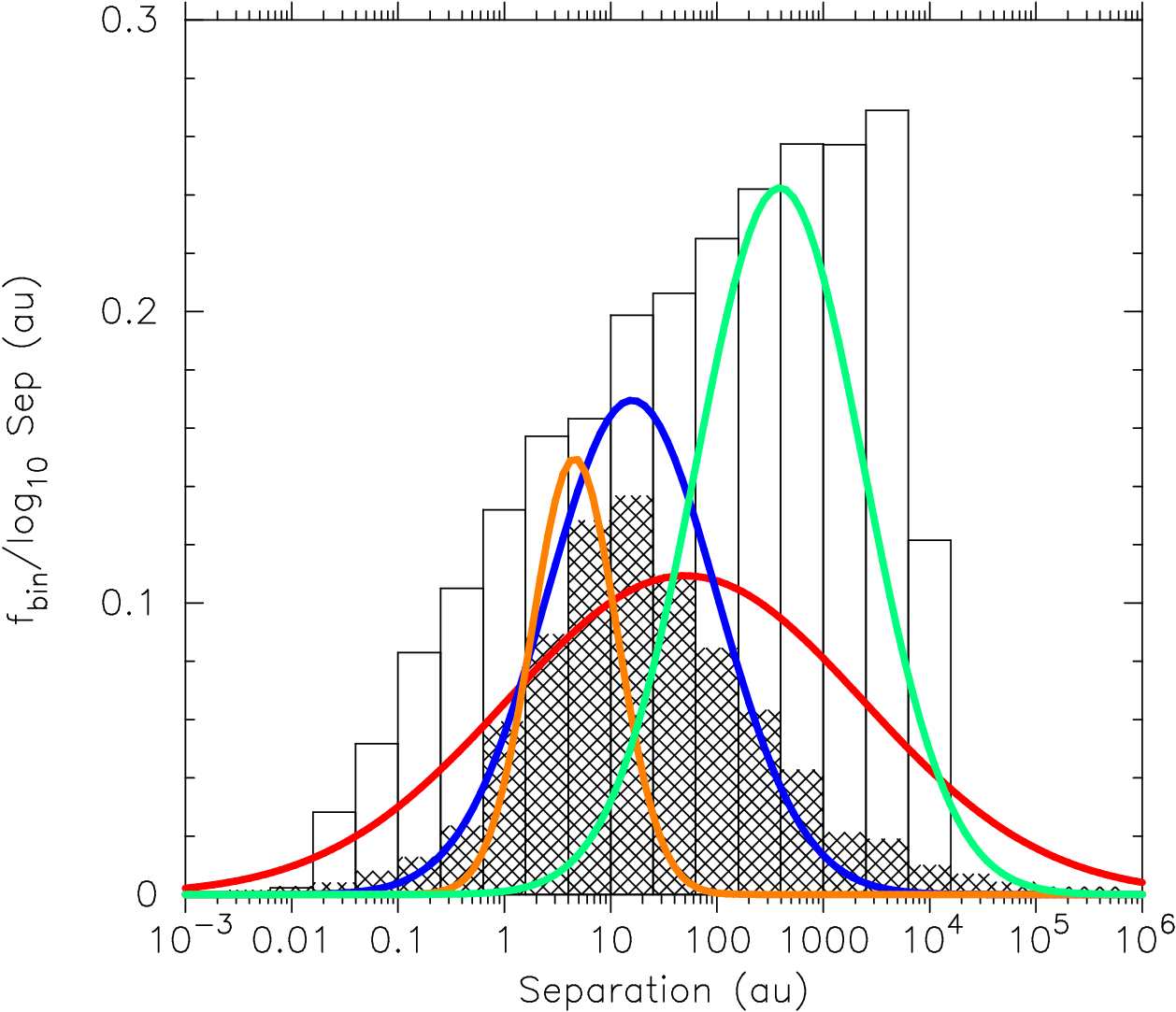}}}
\caption[bf]{The binary properties adopted in our simulations. Panel (a) shows the binary fraction as a function of primary mass for the field binaries (the coloured symbols and lines), and the mass-independent binary fraction from the \citet{Kroupa95a} population (the grey line). Panel (b) shows the various log-normal fits to binaries in the Galactic field (again, this is a strong function of primary mass) and the cross-hatched histogram shows the resultant separation distribution (which is dominated by the most numerous M-dwarf systems). The open histogram is the \citet{Kroupa95a} separation distribution. In both panels the orange, blue, red, green and magenta symbols/lines are for the field brown dwarf, M-dwarf, G-dwarf, A-star and OB star binaries, respectively.} 
\label{binary_props}
  \end{center}
\end{figure*}

In Fig.~\ref{binary_props-a} we show the binary fractions as a function of primary mass for both the Field-like and \citet{Kroupa95a} binary populations. In Fig.~\ref{binary_props-b} we show the log-normal fits to the Field binary separation distributions (which are a function of the primary mass star). The resultant Field-like separation distribution from sampling the IMF is shown by the cross-hatched histogram. The \citet{Kroupa95a} separation distribution is shown by the open histogram.

\begin{table*}
\caption[bf]{Binary properties of systems set up with the distributions observed in the Galactic field. We show the spectral type of the primary mass, the main sequence mass range this corresponds to, the binary fraction $f_{\rm bin}$, and for stars less massive than 3\,M$_\odot$ we show the mean separation $\bar{a}$, and the mean (${\rm log}\,\bar{a}$) and variance ($\sigma_{{\rm log}\,\bar{a}}$) of the log-normal fits to these distributions. OB stars ($m_p>$3.0\,M$_\odot$) are not drawn from a log-normal distribution but instead are drawn from a log-uniform \citet{Opik24} distribution in the range 0 - 50\,au.}
\begin{center}
\begin{tabular}{|c|c|c|c|c|c|c|}
\hline 
Type & Primary mass & $f_{\rm bin}$ & $\bar{a}$ & ${\rm log}\,\bar{a}$ & $\sigma_{{\rm log}\,\bar{a}}$ & Ref. \\
\hline
Brown dwarf &  $0.02 < m_p/$M$_\odot \leq 0.08$ & 0.15 & 4.6\,au & 0.66 & 0.4 & \citet{Burgasser07,Thies07} \\
\hline
M-dwarf & $0.08 < m_p/$M$_\odot \leq 0.45$ & 0.34 & 16\,au & 1.20 & 0.80 & \citet{Bergfors10,Janson12} \\
\hline 
F, G, K & $0.8 < m_p/$M$_\odot \leq 1.2$ & 0.46 & 50\,au & 1.70 & 1.68 & \citet{Raghavan10} \\
\hline
A & $1.5 < m_p/$M$_\odot \leq 3.0$ & 0.48 & 389\,au & 2.59 & 0.79 & \citet{DeRosa14} \\
\hline
OB & $m_p>$3.0\,M$_\odot$ & 1.00 & {\"O}pik & 0 -- 50\,au & log-uniform & \citet{Sana13} \\ 
\hline
\end{tabular}
\end{center}
\label{field_props}
\end{table*}

\begin{table*}
\caption[bf]{Summary of simulation set-ups. The columns show the simulation suite number, number of stars,  $N_{\rm stars}$, initial virial ratio of the regions, $\alpha_{\rm vir}$, fractal dimension of the region, $D$, the median local stellar density $\tilde{\rho}$ and the initial binary population.}
\begin{center}
\begin{tabular}{|c|c|c|c|c|c|c|}
\hline 
Sim.\,No. & $N_{\rm stars}$ & $\alpha_{\rm vir}$& $D$ & $\tilde{\rho}$ & Binary population \\
\hline
1 & 1500 & 0.3 & 1.6 & 10\,000\,M$_\odot$\,pc$^{-3}$ & Galactic field \\
2 & 1500 & 0.3 & 3.0 & 100\,M$_\odot$\,pc$^{-3}$ & Galactic field \\
\hline
3 & 1500 & 0.3 & 1.6 & 10\,000\,M$_\odot$\,pc$^{-3}$ & \citet{Kroupa95a} `universal' \\
\hline
\end{tabular}
\end{center}
\label{cluster_sims}
\end{table*}

\subsection{Dynamical evolution}

We place each binary at the centre of mass of the stellar systems in the fractal distribution. We use the \texttt{kira} integrator \citep{Zwart99,Zwart01} to perform the $N$-body integrations and we evolve our star-forming regions for 10\,Myr. We do not include stellar evolution in the simulations. The different simulations are summarised in Table~\ref{cluster_sims}.

\subsection{Stability of planetary systems}

%Check the HW criteria for individual systems. 

We define whether a planet can reside on a stable orbit in or around a binary star system using the results of \citet{Holman99}, who conducted numerical simulations of mass-less test particles within binary star systems to obtain a so-called critical semimajor axis. \citet{Holman99} obtained a polynomial fit to their results that depends on both the eccentricity, mass ratio and semimajor axis of the binary. 

The following stability equations are reliable in most cases, but there are systems that could be stable according to the \citet{Holman99} criteria when in reality they are not, and vice versa. There is an extensive literature that describes modifications and alternatives to the \citet{Holman99} criteria \citep[e.g.][]{PilatLohinger02,PilatLohinger03,Lam18,Quarles18,Quarles20}, but we adopt these criteria merely as a rough indication of whether a planet could remain stable on either a circumprimary or circumbinary orbit.

For an S-type orbit, where the planet orbits one of the stars in the binary system, the critical semimajor axis $a_{cS}$, is the maximum distance from the planet hosting star in which a planet can reside on a stable orbit and is given by
\begin{equation}
\begin{array}{lll} a_{cS} & = & [0.464 - 0.38\mu - 0.631e \,\,+ \vspace*{0.1cm} \\  &  & 0.586\mu e + 0.15e^2 - 0.198\mu e^2]a_{\rm bin}  \end{array}
\label{stab_crit_S}
\end{equation}
where $e$ is the eccentricity of the binary, $a_{\rm bin}$ is the semimajor axis of the binary and 
\begin{equation}
\mu =  \frac{m_s}{m_p + m_s},
\label{reduced_mass}
\end{equation}
where $m_p$ and $m_s$ are the masses of the primary and secondary component stars of the binary, respectively. Inspection of Eqn.~\ref{stab_crit_S} immediately shows that for a circular ($e = 0$), equal-mass binary ($m_p = m_s$, $\mu = 0.5$), the critical semimajor axis would be $a_{cS} = 0.27a_{\rm bin}$, whereas a more eccentric (e.g $e = 0.5$) binary would have a significantly smaller critical semimajor axis ($a_{cS} = 0.12a_{\rm bin}$), thus reducing the orbital parameter space in which a planet could reside on a stable orbit. 

Throughout the paper we only consider the stability of planets orbiting the primary star. However, Eqn.~\ref{stab_crit_S} could also be applied to planets orbiting the secondary star, with the proviso that Eqn.~\ref{reduced_mass} is changed to:
 \begin{equation}
\mu_s =  \frac{m_p}{m_p + m_s}.
\end{equation}
In tests we find that the fraction of stable \emph{circumsecondary} planets at a given separation will be lower than for planets orbiting the primary.

For a P-type orbit, where the planet orbits the entire binary star system, the critical semimajor axis $a_{cP}$, is the minimum distance from the binary system on which a planet can reside on a stable orbit and is given by
\begin{equation}
\begin{array}{lll} a_{cP} & = & [1.6 + 5.1e - 2.22e^2 + 4.12\mu \,\,- \vspace*{0.1cm} \\  &  & 4.27e\mu - 5.09\mu^2 + 4.61e^2\mu^2]a_{\rm bin} \end{array}
\label{stab_crit_P}
\end{equation}
where the symbols are the same as in Eqn.~\ref{stab_crit_S}. For circumbinary orbits, a circular, equal-mass binary would have a critical semimajor axis $a_{cP} = 2.39a_{\rm bin}$, whereas a more eccentric ($e = 0.5$) binary would have $a_{cP} = 3.60a_{\rm bin}$. As for the circumprimary case, the more eccentric the binary, the smaller the parameter space is in which a planet could reside.  

These analytic formulae provide a reasonable estimate of the stability (or otherwise) of a planet on a given orbit. However, there are exceptions in the literature and we refer the interested reader to \citet{Parker13c} and \citet{Lam18,Quarles18,Kong21} for further details. For the remainder of this paper, we will assume these formulae are valid for all systems, but we recommend any assessment of the stability of an observed binary system should use additional/alternative stability criteria. 

\subsection{von Zeipel-Lidov-Kozai mechanism}

The von Zeipel-Lidov-Kozai (vZLK) mechanism \citep{vonZeipel10,Lidov62,Kozai62} describes the transfer of angular momentum onto the least massive object in a three-body system, following a change of more than $39^\circ$  but less than $141^\circ$ in the mutual inclination between the two more massive bodies. Previous work \citep{Innanen97,Fabrycky07,Malmberg07a,Parker09c} has shown that the vZLK mechanism could occur in stellar binary systems due to interactions in the natal star-forming region. We determine the change (if any) in the inclination of the binary systems and assume that a planet(s) could subsequently be  subjected to the vZLK mechanism if the change in inclination of the binary is between $39^\circ$ and $141^\circ$. 

%\subsection{$N$-body simulations}

\subsection{Calculating the habitable zones in binaries}

We also calculate any change in the habitable zone around binary systems, following the method outlined in \citet{Wootton19}. This involves defining an isophote, i.e.\,\,a region around each star where the temperature is such that water could exist in liquid form. Occasionally, changes to the orbit of the binary as a result of dynamical interactions can cause the two stars to move closer together, increasing the size of the isophote-based habitable zone. In \citet{Wootton19} we performed this calculation for the binaries in the first suite of our $N$-body simulations (field binaries in a highly substructured [$D=1.6$] region) and for comparison we will extend this analysis to the smooth ($D = 3.0$) star-forming regions and the simulations with the \citet{Kroupa95a} `universal' initial binary population. 

The boundaries of the habitable zone around a binary are calculated using the following equation:
\begin{equation}
l_{\rm x-bin} = \left[\frac{W_pL_p}{L_\odot/l_{\rm x-\odot}^2 - W_sL_s/r_{\rm pl-s}^2} \right]^{1/2},
\end{equation} 
where $L_p$ and $L_s$ are the luminosities of the primary and secondary stars, respectively, and are calculated from the masses of the stars using the standard
\begin{equation}
L_i/{\rm L_\odot} = \left(M_i/{\rm M_\odot} \right)^{3.5},
\end{equation}
where the subscript $_i$ refers to either the primary component $_p$ or the secondary component $_s$. $r_{\rm pl-s}$ is the distance from a hypothetical planet to the secondary star in order to generate a reference position from which $l_{\rm x-bin}$ is calculated \citep{Kaltenegger13,Haghighipour13}. $W_p$ and $W_s$ are the so-called spectral weight factors, which determine the amount of flux contributed by the primary and secondary stars, and are the function of the effective temperature $T_i$ (normalised to the Sun's effective temperature, see below) and the cloud cover $f$:
\begin{equation}
W_i(f, T_i) = \left[1 + \alpha_{\rm x}(T_i)l_{\rm x-\odot}^2 \right]^{-1}.
\end{equation}
$\alpha_{\rm x}(T_i)$ is given by
\begin{equation}
\alpha_{\rm x}(T_i) = a_{\rm x}T_i + b_{\rm x}T_i^2 + c_{\rm x}T_i^3 + d_{\rm x}T_i^4 ,
\end{equation}
where
\begin{equation}
T_i({\rm K}) = T_{\rm Star}({\rm K}) - 5780,
\end{equation}
and  $a_{\rm x}$, $b_{\rm x}$, $c_{\rm x}$  and $d_{\rm x}$ are coefficients which depend on the type of habitable zone under consideration and are found in \citet{Kopparapu13b} and also summarised in table~1 in \citet{Kaltenegger13}. 

The only other variable yet to be defined is $l_{\rm x-\odot}$, which represents the boundaries of the habitable zone for the Sun. Typically, calculations determine a narrow habitable zone with values between $l_{\rm x-\odot} = 0.97$\,au and $l_{\rm x-\odot} = 1.67$\,au \citep{Kopparapu13b} and an extended empirical habitable zone between $l_{\rm x-\odot} = 0.75$\,au and $l_{\rm x-\odot} = 1.77$\,au. We elect to consider the change only to the narrow habitable zone, and also note that other authors apply a much more stringent definition to what constitutes `habitable', including the constraint that any planets would not be destabilised by the orbital motion of the binary \citep[e.g][]{Georgakarakos19}. However, for the purposes of this paper we merely wish to quantify how many binary systems could have enlarged isophote-based habitable zones as a result of dynamical hardening or softening encounters on the binary.

\section{Results}
\label{results}

\subsection{Change in binary properties}

\begin{figure*}
 \begin{center}
\setlength{\subfigcapskip}{10pt}
\hspace*{-1.5cm}\subfigure[Field binaries, $D = 1.6$]{\label{binary_change-a}\rotatebox{270}{\includegraphics[scale=0.27]{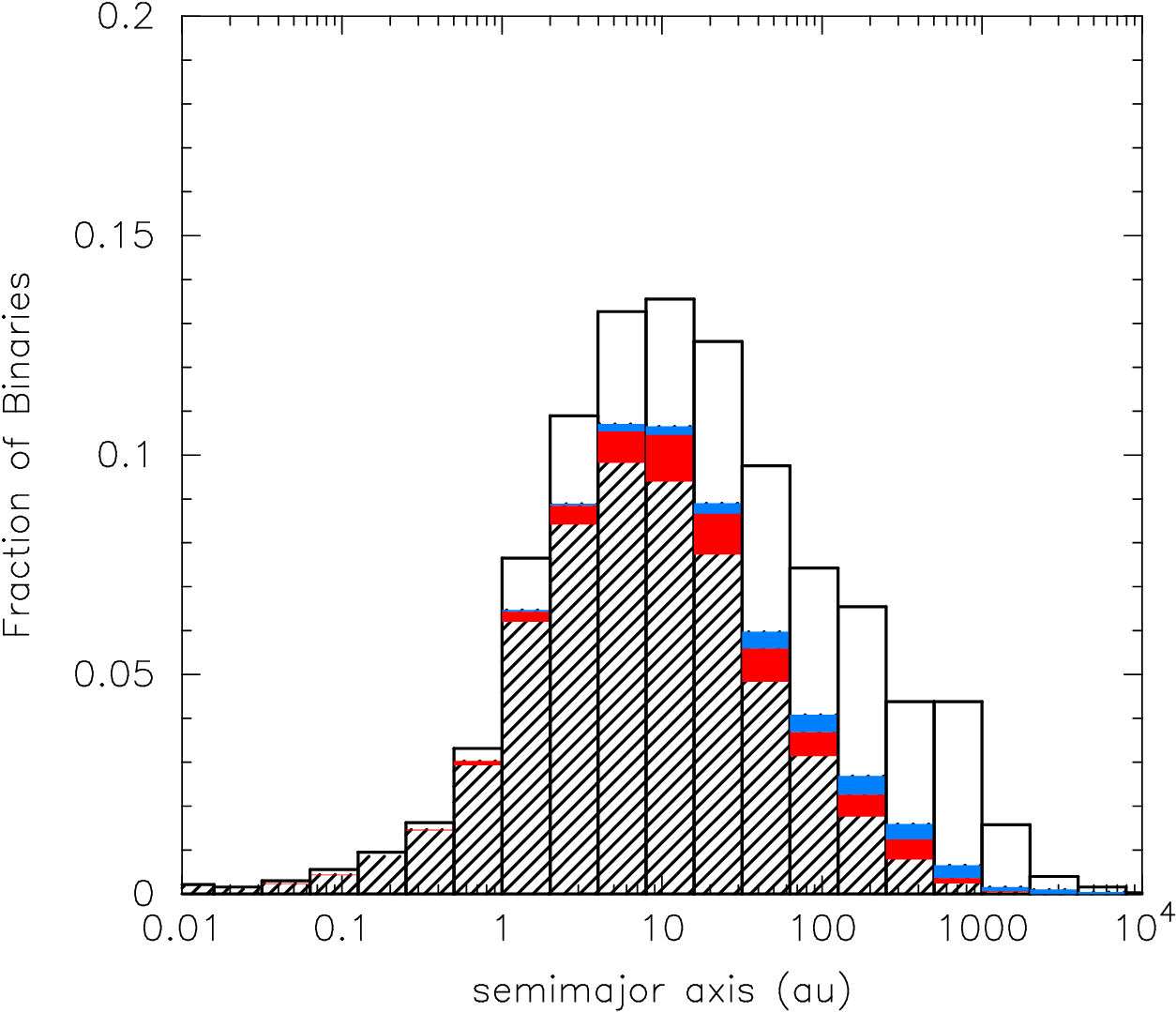}}}
\hspace*{0.5cm}\subfigure[Field binaries, $D = 3.0$]{\label{binary_change-b}\rotatebox{270}{\includegraphics[scale=0.27]{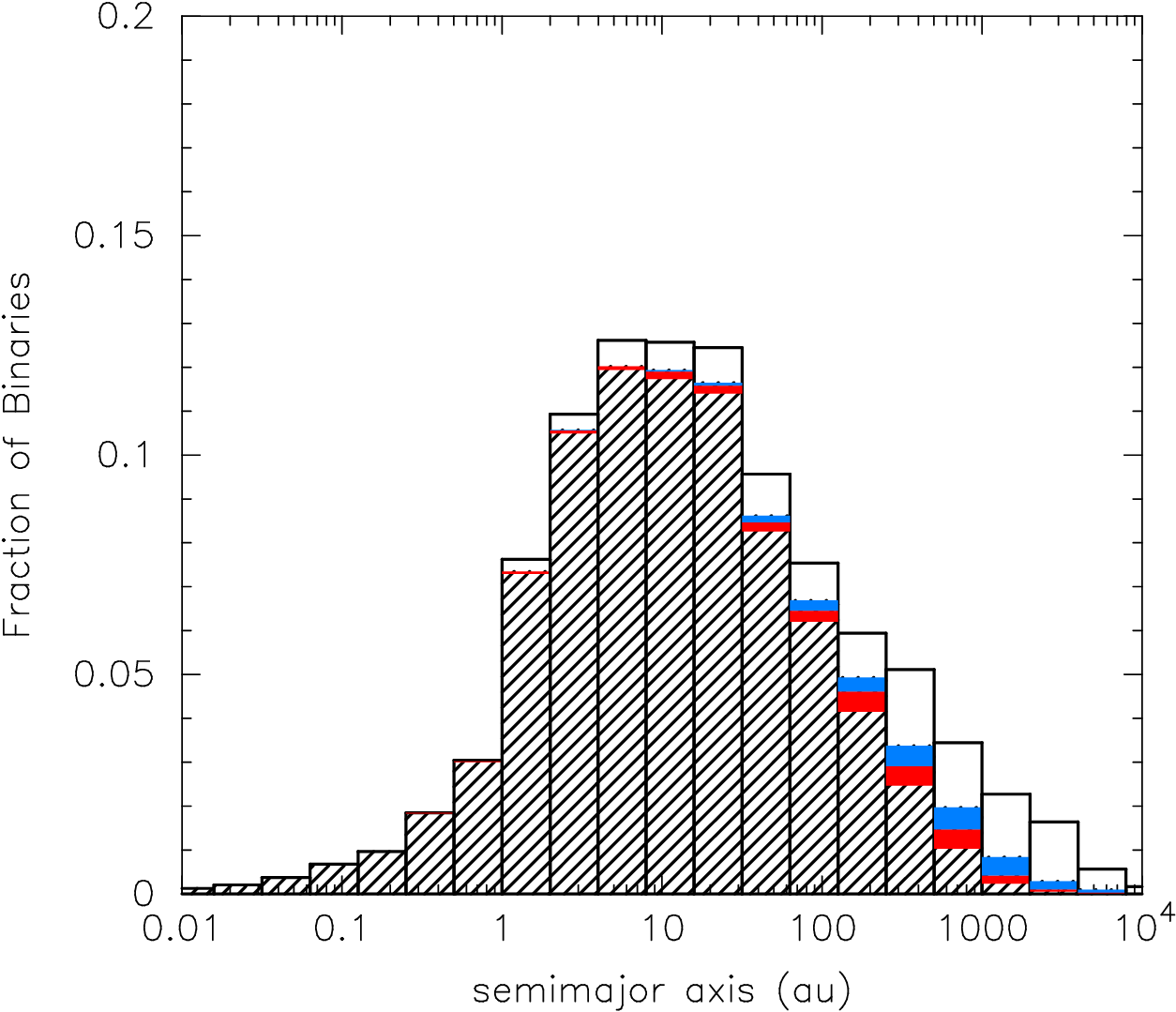}}}
\hspace*{0.5cm}\subfigure[\citet{Kroupa95a} binaries, $D = 1.6$]{\label{binary_change-c}\rotatebox{270}{\includegraphics[scale=0.27]{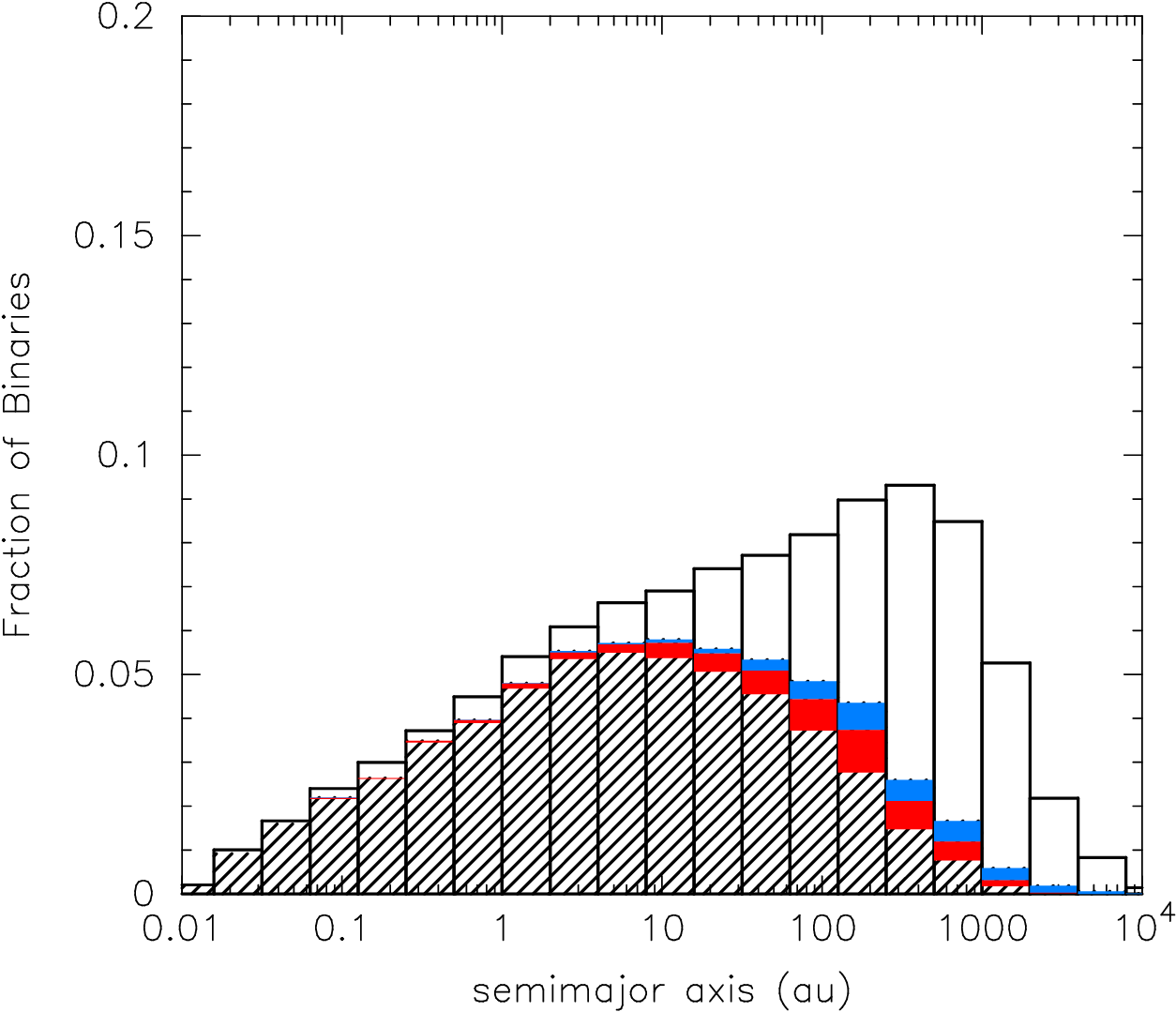}}}
\caption[bf]{Change in semimajor axes distributions for different initial conditions. Panels (a) and (b) show populations of binaries with semimajor axes drawn from the distributions observed in the Galactic field, and panel (c) shows the \citet{Kroupa95a} binary population. The degree of substructure (fractal dimension $D$) is indicated. The open histogram shows the initial distribution, and the hashed histogram shows the distribution after 10\,Myr of evolution. Binaries that have their semimajor axes decreased (hardened) by more than 10\,per cent of their original semimajor axis are shown by the red histograms, and binaries whose semimajor axes increase (softened) by more than 10\,per cent are shown by the blue histograms.} 
\label{binary_change}
 \end{center}
\end{figure*}

The effects of a dense stellar environment on binary systems has been well-documented in the literature \citep[e.g.][]{Heggie75,Hills75a,Kroupa95a,Kroupa95b,Fregeau06,Parker11c}. There are multiple possible outcomes of a dynamical encounter on a binary system, including total ionisation, or a change in the orbital parameters of the system. Binaries with a high binding energy relative to their surroundings are often termed dynamically `hard' systems, and an encounter will likely increase their energy by reducing the semimajor axis, whereas binaries with a smaller energy relative to their surroundings are `soft' and an encounter is likely to further reduce their energy, increasing the semimajor axis \citep{Heggie75,Hills75a,Hills75b}. 

However, the binaries that are most interesting from the point of view of planetary stability are so-called intermediate systems, i.e. those that are neither hard nor soft, and have a binding energy comparable to the surroundings. In the dense environments we model here, the semimajor axis of intermediate binaries can both increase and decrease, and we demonstrate this in Fig.~\ref{binary_change}. 

In this figure, we show the initial semimajor axis distribution by the open histograms, and the final distributions (after 10\,Myr of dynamical processing) by the hashed histograms. Within the final distributions, we show the proportion of binaries whose semimajor axis decreases by $\geq$10\,per cent of the initial value by the solid red shading, and the proportion of binaries whose semimajor axis increases by $\geq$10\,per cent of the initial value by the solid blue shading. 

In panel (a) of this figure, we show the result for our highly substructured simulations ($D = 1.6$) which contain a population of Field-like binaries. %We plot the  change in semimajor axis against the initial semimajor axis, with systems whose semimajor axis increases over 10\,Myr shown in blue, and systems whose semimajor axis decreases shown in red. 

In panel (b) we show the evolution of the same binary population, but this time in the star-forming regions with no initial substructure ($D = 3.0$). Overall, more binaries are significantly disrupted in the substructured star-forming regions, and binaries with smaller semimajor axes are also susceptible. Although these simulations contain the same number of stars and have the same radii (1\,pc), the more substructured regions have a higher median local stellar density, and also undergo a more violent relaxation \citep{Allison10}, thus dynamically processing the binary population to a greater extent. 

Finally, in panel (c) of this figure we show the evolution of the \citet{Kroupa95a} binary population, within  highly substructured simulations ($D = 1.6$). These simulations also destroy a significant fraction of binaries, but these tend to be very wide/soft  ($>$100\,au) binaries.

It is the intermediate binary systems (separations between 10 -- 100\,au), and the change to their orbital parameters, that could affect the fraction of stable planetary systems.

\subsection{Stability of planets in and around binaries}

We now calculate the fraction of binary systems that could host stable hypothetical planets at different distances from either the primary star (for circumprimary or S-type planets)  or the whole binary (circumbinary, or P-type planets). 

\subsubsection{Circumprimary planets}

\begin{figure*}
  \begin{center}
\setlength{\subfigcapskip}{10pt}
%\hspace*{0.3cm}\subfigure[$D = 1.6$]{\label{}\rotatebox{270}{\includegraphics[scale=0.35]{plot_HW_S_all_Or_C0p3F1p61pB_F10.ps}}}
%\hspace*{0.5cm}\subfigure[$D = 3.0$]{\label{}\rotatebox{270}{\includegraphics[scale=0.35]{plot_HW_S_all_Or_C0p3F3p01pB_F10.ps}}}
\hspace*{0.3cm}\subfigure[$D = 1.6$]{\label{circum_prime_fracs_all-a}\rotatebox{270}{\includegraphics[scale=0.35]{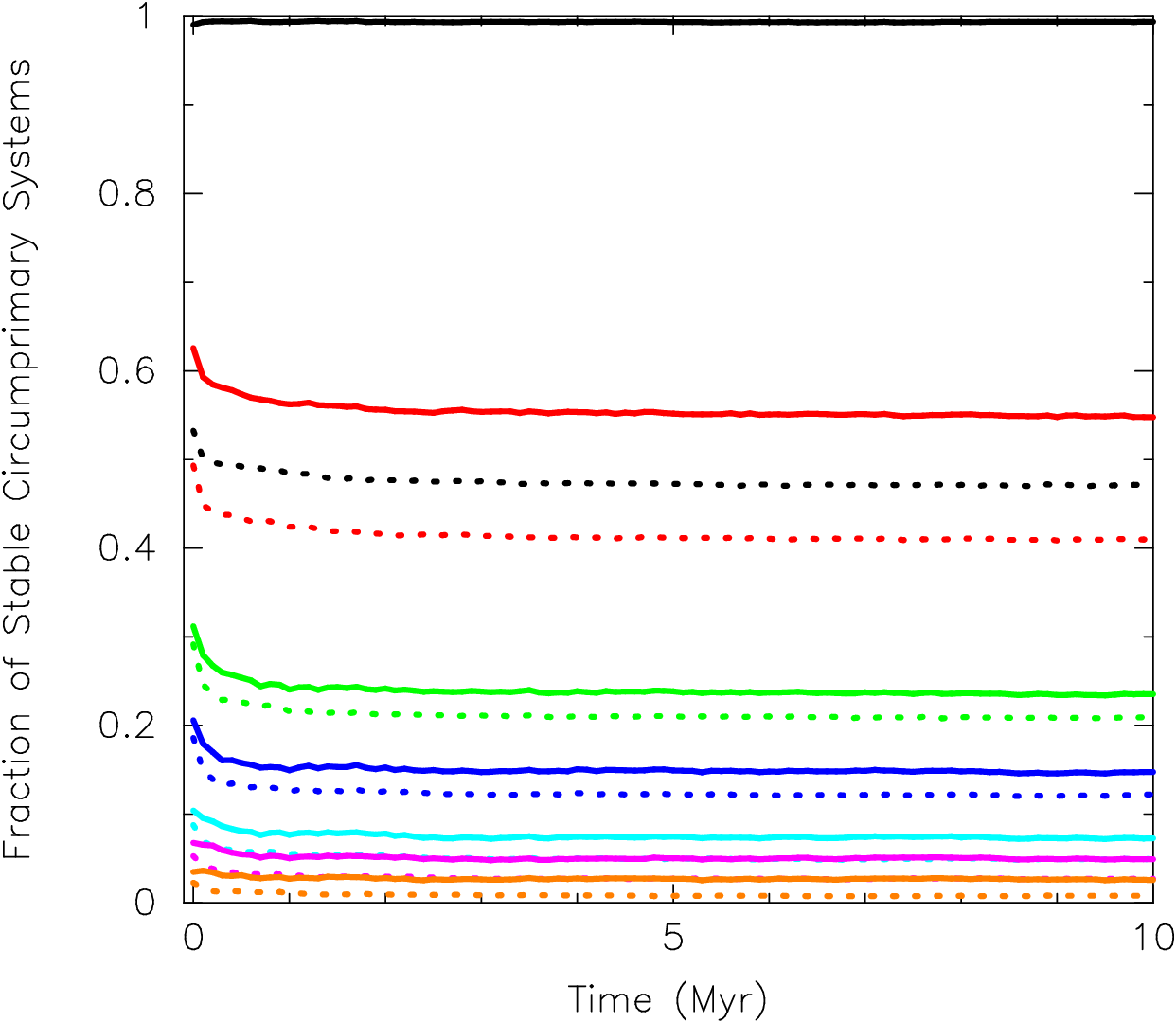}}}
\hspace*{0.5cm}\subfigure[$D = 3.0$]{\label{circum_prime_fracs_all-b}\rotatebox{270}{\includegraphics[scale=0.35]{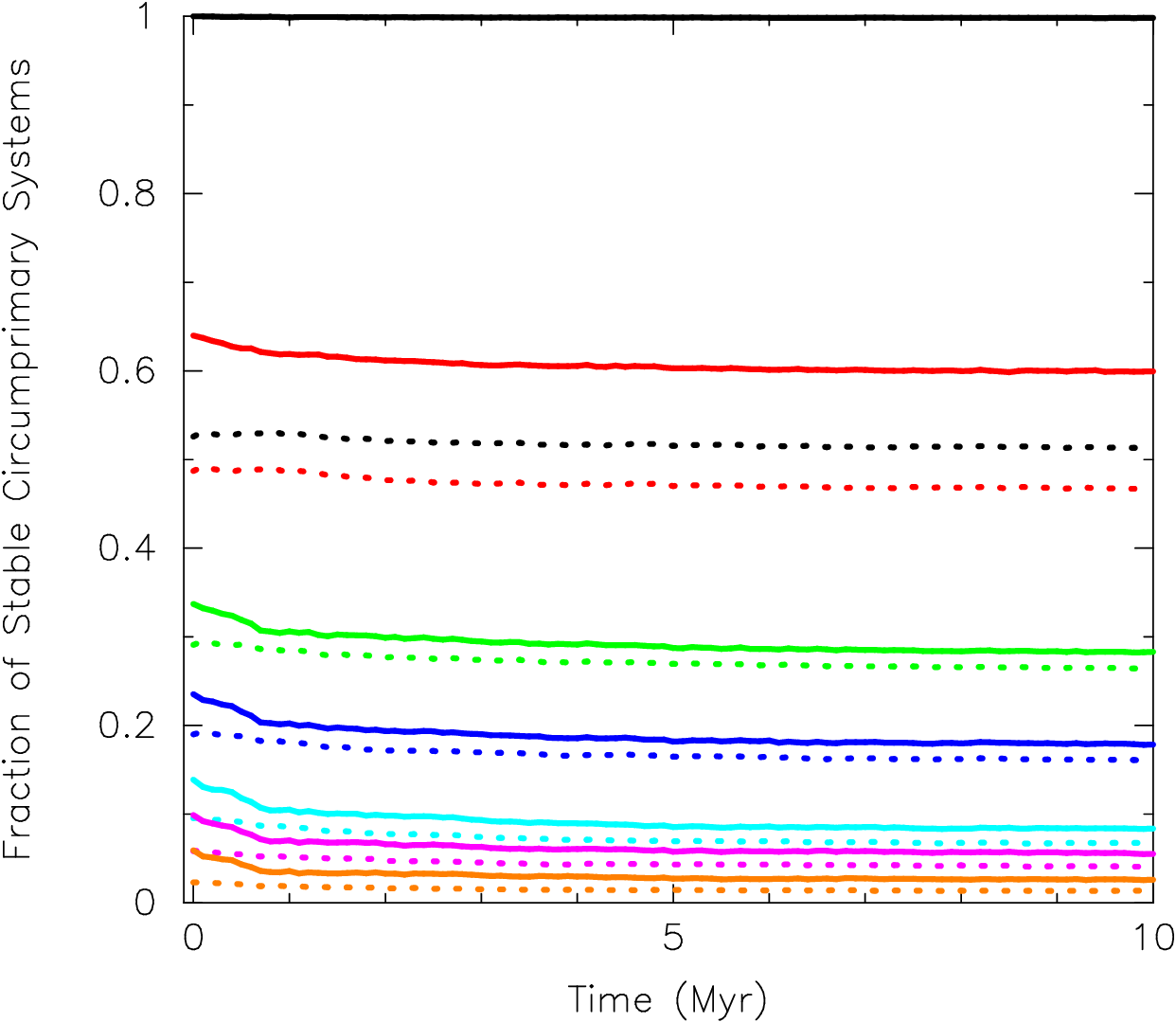}}}
\caption[bf]{Fraction of \emph{all} binary systems in which planets on S-type (circumprimary) orbits are stable according to the \citet{Holman99} criteria. The solid lines show the fraction of all binary systems that could host a stable planet, whereas the dotted lines show the corresponding fractions of binary systems that have initial semimajor axes in the range 10--1000\,au that could host stable planets. The black lines indicate the fraction of binary systems that could host a planet within some distance from the primary star (and as such is nearly unity when we do not apply a cut to the binary semimajor axes); the red lines indicate the fraction of binaries that could host a stable planet at 1\,au from the primary star; the green lines indicate the fraction of binaries that could host a planet at 5\,au; the dark blue lines indicate the fraction of binaries that could host a planet at 10\,au; the cyan lines indicate the fraction of binaries that could host a planet at 30\,au; the magenta lines indicate the fraction of binaries that could host a planet at 50\,au and the orange lines indicate the fraction of binaries that could host a planet at 100\,au from the primary star. In panel (a) we show the results for a evolution of a population of field-like binaries in a highly substructured (fractal dimension $D = 1.6$) star-forming region, and in panel (b) we show the results for a star-forming region with no substructure, i.e.\,\,smooth (fractal dimension $D = 3.0$).} 
\label{circum_prime_fracs_all}
  \end{center}
\end{figure*}

In Fig.~\ref{circum_prime_fracs_all} we show the fraction of all binaries that could host a stable planet over time. Here, we divide the number of binaries   that could host a planet at a given distance by the total number of initial binaries. This leads to a slight drop in the fractions of stable planet-hosting binaries due to the dynamical destruction of some of the primordial binaries.

The solid lines in Fig.~\ref{circum_prime_fracs_all} are the fractions of binaries of any semimajor axis that could host stable planets, whereas the dotted lines are the fractions of binaries with semimajor axes in the range 10 -- 1000\,au, which corresponds to the typical separation range that observations of binary systems in star-forming regions are sensitive to \citep[e.g][]{Kohler98,Duchene99,Ratzka05,Patience02,King12a,Duchene13b,Duchene18}. The coloured lines correspond to the fraction of binaries that could host stable planets at different distances from the host star; the red lines are for planets at 1\,au, the green lines are for planets at 5\,au, the dark blue lines are for planets at 10\,au, the cyan lines are for planets at 30\,au, the magenta lines are for planets at 50\,au and the orange lines are for planets at 100\,au. The black lines are binaries that could host a planet at some distance, i.e. there are very few binaries that are so eccentric that they couldn't host any stable planets.

For these circumprimary planets, the larger the distance of the planet from the primary star, the less likely the planet is to be stable. Furthermore, if we restrict our analysis to binaries with semimajor axes in the observed range in star-forming regions (10 -- 1000\,au) we see that the fraction of binaries that could host stable planets is even lower. There is a small dependence on the initial conditions of the star-forming regions; regions with initial substructure contain fewer binaries that could host stable planets (Fig.~\ref{circum_prime_fracs_all-a}) than regions with no initial substructure (Fig.~\ref{circum_prime_fracs_all-b}), though the differences are  of order 5\,per cent.

\subsubsection{Circumbinary planets}

\begin{figure*}
  \begin{center}
\setlength{\subfigcapskip}{10pt}
%\hspace*{0.3cm}\subfigure[$D = 1.6$]{\label{}\rotatebox{270}{\includegraphics[scale=0.35]{plot_HW_P_all_Or_C0p3F1p61pB_F10.ps}}}
%\hspace*{0.5cm}\subfigure[$D = 3.0$]{\label{}\rotatebox{270}{\includegraphics[scale=0.35]{plot_HW_P_all_Or_C0p3F3p01pB_F10.ps}}}
\hspace*{0.3cm}\subfigure[$D = 1.6$]{\label{circum_bin_fracs_all-a}\rotatebox{270}{\includegraphics[scale=0.35]{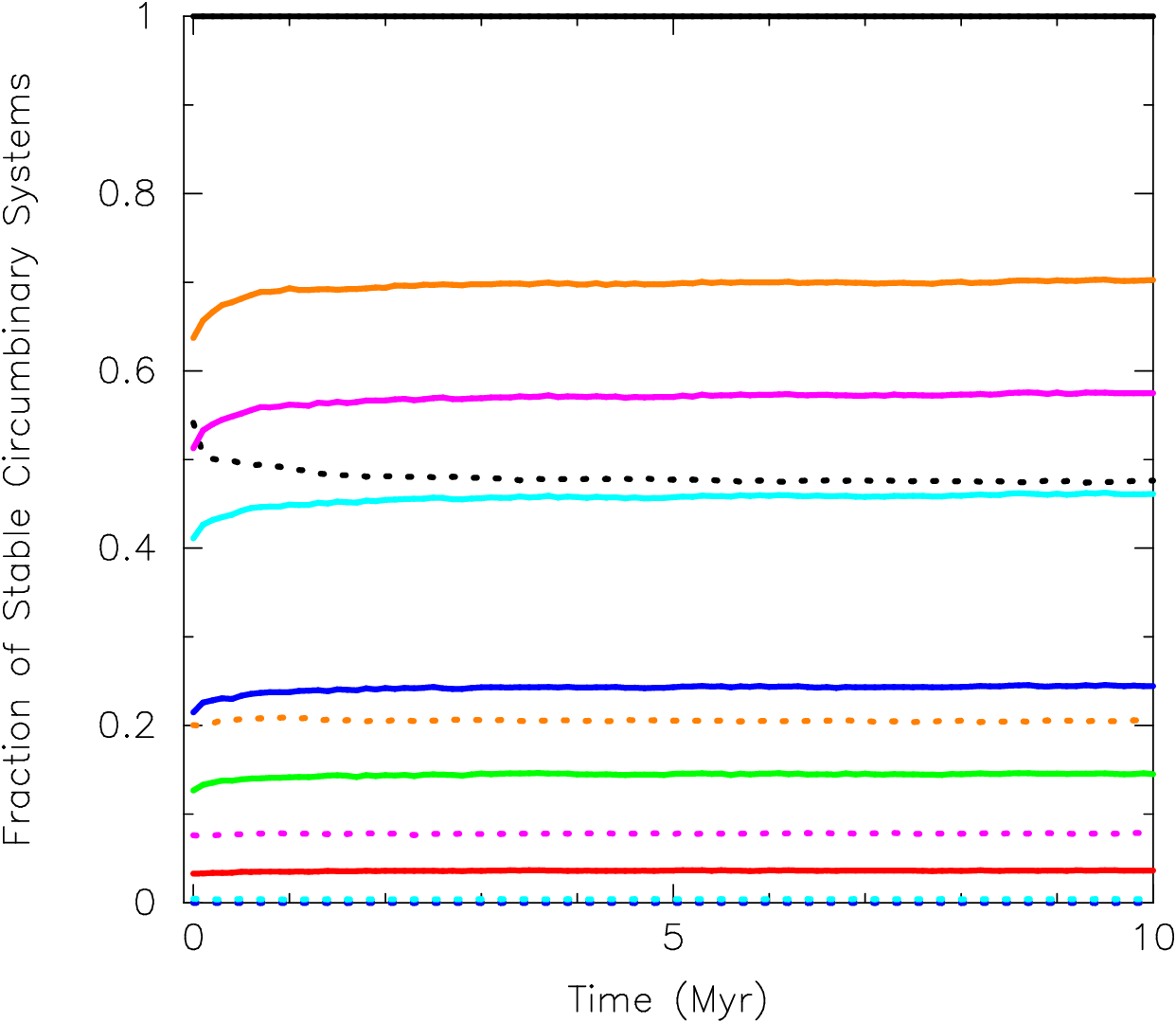}}}
\hspace*{0.5cm}\subfigure[$D = 3.0$]{\label{circum_bin_fracs_all-b}\rotatebox{270}{\includegraphics[scale=0.35]{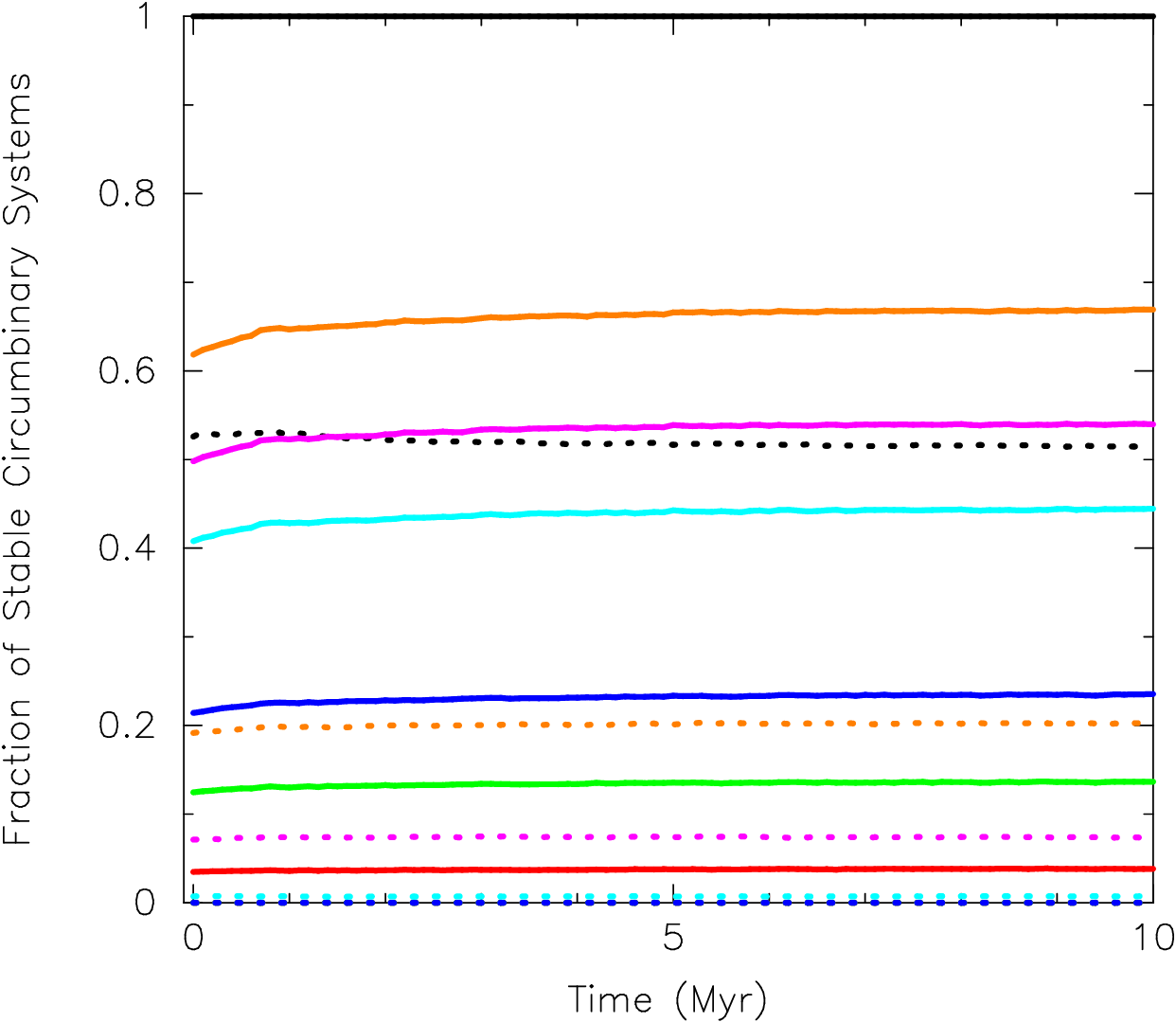}}}
\caption[bf]{Fraction of \emph{all} binary systems in which planets on P-type (circumbinary) orbits are stable according to the \citet{Holman99} criteria. The solid lines show the fraction of all binary systems that could host a stable planet, whereas the dotted lines show the corresponding fractions of binary systems that have initial semimajor axes in the range 10--1000\,au that could host stable planets. The black lines indicate the fraction of binary systems that could host a planet within some distance from the binary (and as such is nearly unity when we do not apply a cut to the binary semimajor axes); the red lines indicate the fraction of binaries that could host a stable planet at 1\,au from the binary; the green lines indicate the fraction of binaries that could host a planet at 5\,au; the dark blue lines indicate the fraction of binaries that could host a planet at 10\,au; the cyan lines indicate the fraction of binaries that could host a planet at 30\,au; the magenta lines indicate the fraction of binaries that could host a planet at 50\,au and the orange lines indicate the fraction of binaries that could host a planet at 100\,au from the binary system. In panel (a) we show the results for a evolution of a population of field-like binaries in a highly substructured (fractal dimension $D = 1.6$) star-forming region, and in panel (b) we show the results for a star-forming region with no substructure, i.e.\,\,smooth (fractal dimension $D = 3.0$).} 
\label{circum_bin_fracs_all}
  \end{center}
\end{figure*}

We show the stable fractions of binaries with circumbinary (P-type) planets at various distances from the binary in Fig.~\ref{circum_bin_fracs_all}. As in Fig.~\ref{circum_prime_fracs_all}, the solid lines are the fraction of stable binaries at all semimajor axes, whereas the dotted lines are for binaries with semimajor axes in the range 10 -- 1000\,au. The black lines are the fraction of binaries that could host a planet at some distance, whereas the red, green, dark blue, cyan, magenta and orange lines are the fractions of binaries that could host stable planets at 1, 5, 10, 30, 50 and 100\,au, respectively. 

As one might expect, the fractions of binaries that could host stable circumbinary planets are flipped compared to the fractions of binaries that could host circumprimary planets; the further the planetary distance, the higher the probability of that planet being stable. This is in part due to the demographics of the binary population we adopt as the initial conditions; more than half of binaries in the Galactic field have semimajor axes less than 50\,au, so the fraction of binaries that could host stable planets at smaller orbital distances will decrease.  For planets at specified distances, the fraction of stable binaries increases slightly over time, which is likely due to the destruction of wide binaries that would be unlikely to host a planet at any distance.

When we consider only binaries in the observed range (10 -- 1000\,au), we see that all planets at distances less than 30\,au from the binary would be unstable, and only modest numbers of binaries would be able to host stable planets on more distant orbits. We would therefore expect circumbinary planets to really only be possible on orbits around binaries with semimajor axes less than 10\,au. The black dotted line indicates the fraction of stable binaries in the observed range for a planet at an unspecified distance, and this is the only fraction that decreases over time. This behaviour is due to the binaries in the observed range having their orbits modified (as opposed to being destroyed), which often involves the eccentricity of the binary being excited to values approaching unity. Such high eccentricities would preclude planets on stable orbits within the binary systems. 

As opposed to the circumprimary case, the fraction of stable binary systems is lower in the less substructured star-forming regions (again, by around 5\,per cent, compare panels (a) and (b) in Fig.~\ref{circum_bin_fracs_all}). This is due to the higher number of binaries with wider semimajor axes, which requires the planets to reside further from the binary system in order to be stable.

\subsubsection{Planets in preserved binaries}

We now repeat these calculations but limit ourselves to binary systems that are preserved throughout the duration of the simulation. These are binaries that are present in the initial conditions, and every subsequent snapshot and who retain their birth partners throughout. The behaviour of the stability fractions is very similar to that for all binary systems, and so we show the results in Appendix~\ref{appendix:preserved}.

We show the evolution of the fraction of stable preserved binaries that could host a planet at various distances from the primary star in Fig.~\ref{circum_prime_fracs_pres} and the evolution of the fraction of stable preserved binaries that could host a planet at various distances from the binary in Fig.~\ref{circum_bin_fracs_pres}. 

The striking feature in both figures is the lack of evolution in the fraction of stable binaries. The individual fractions remain more or less constant throughout the 10\,Myr of dynamical evolution, despite, as we have seen in Fig.~\ref{binary_change}, the orbits of some binary systems changing quite drastically. 

The main reason for this lack of evolution is that -- whilst some binaries have their semimajor axes hardened or softened (Fig.~\ref{binary_change}) -- the highest degree of binary processing and destruction tends to be only for binaries with separations in excess of 1000\,au. These binary systems can host a stable planetary system up to 100\,au, but then are broken apart, so do not factor into the fractions of preserved binary systems.

\subsection{The von Zeipel-Lidov-Kozai mechanism}

For the binary systems that are preserved throughout the simulations (i.e.\,\,primordial binary systems whose constituent stars remain bound to their birth partners at every subsequent snapshot in the simulation), we calculate the fraction of binaries whose mutual inclination changes by more than $39^\circ$ but less than $141^\circ$. It is these systems that could undergo the von~Zeipel-Lidov-Kozai mechanism, which can result in the transfer of angular momentum onto any planets within the binary system \citep{Innanen97,Fabrycky07}, under the assumption that the planet(s) orbits(s) are not altered by the interaction that changes the inclination of the binary. 

The result of this process can lead to planet-planet scattering and ejection \citep{Malmberg07a}, and has been shown to be a way of producing Hot Jupiters \citep{Wu03}. Furthermore, the timescale for the vZLK mechanism is of order 1\,Myr for most binaries \citep{Parker09c} and crucially, the timescale is shortest for the outermost planet in the system planet \citep{Kiseleva98,Takeda08}, so if the the outer planet becomes unstable and more eccentric, it is likely to subsequently affect the inner planets.

\begin{figure*}
  \begin{center}
\setlength{\subfigcapskip}{10pt}
\hspace*{0.3cm}\subfigure[$D = 1.6$]{\label{kozai_frac-a}\rotatebox{270}{\includegraphics[scale=0.35]{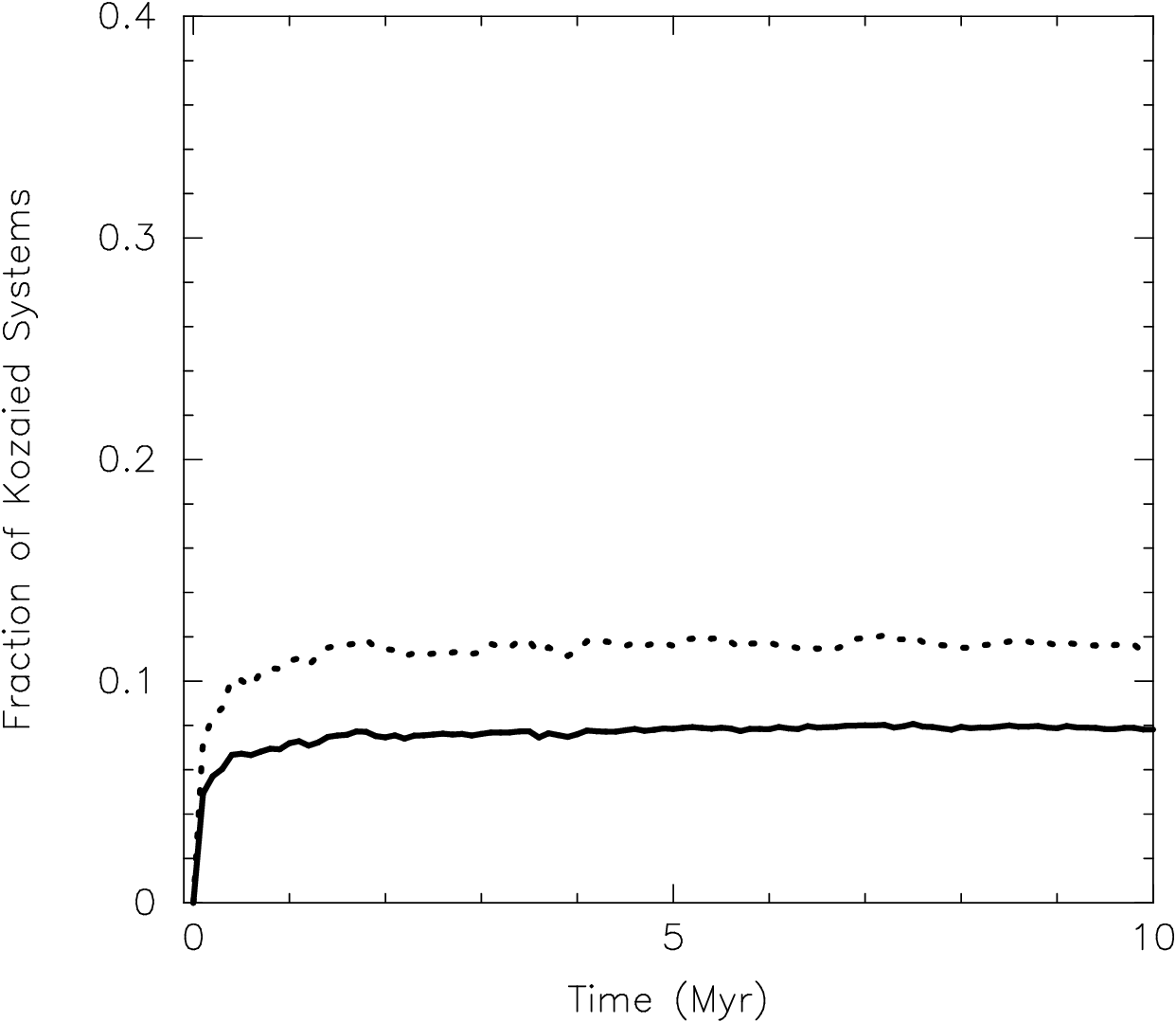}}}
\hspace*{0.5cm}\subfigure[$D = 3.0$]{\label{kozai_frac-b}\rotatebox{270}{\includegraphics[scale=0.35]{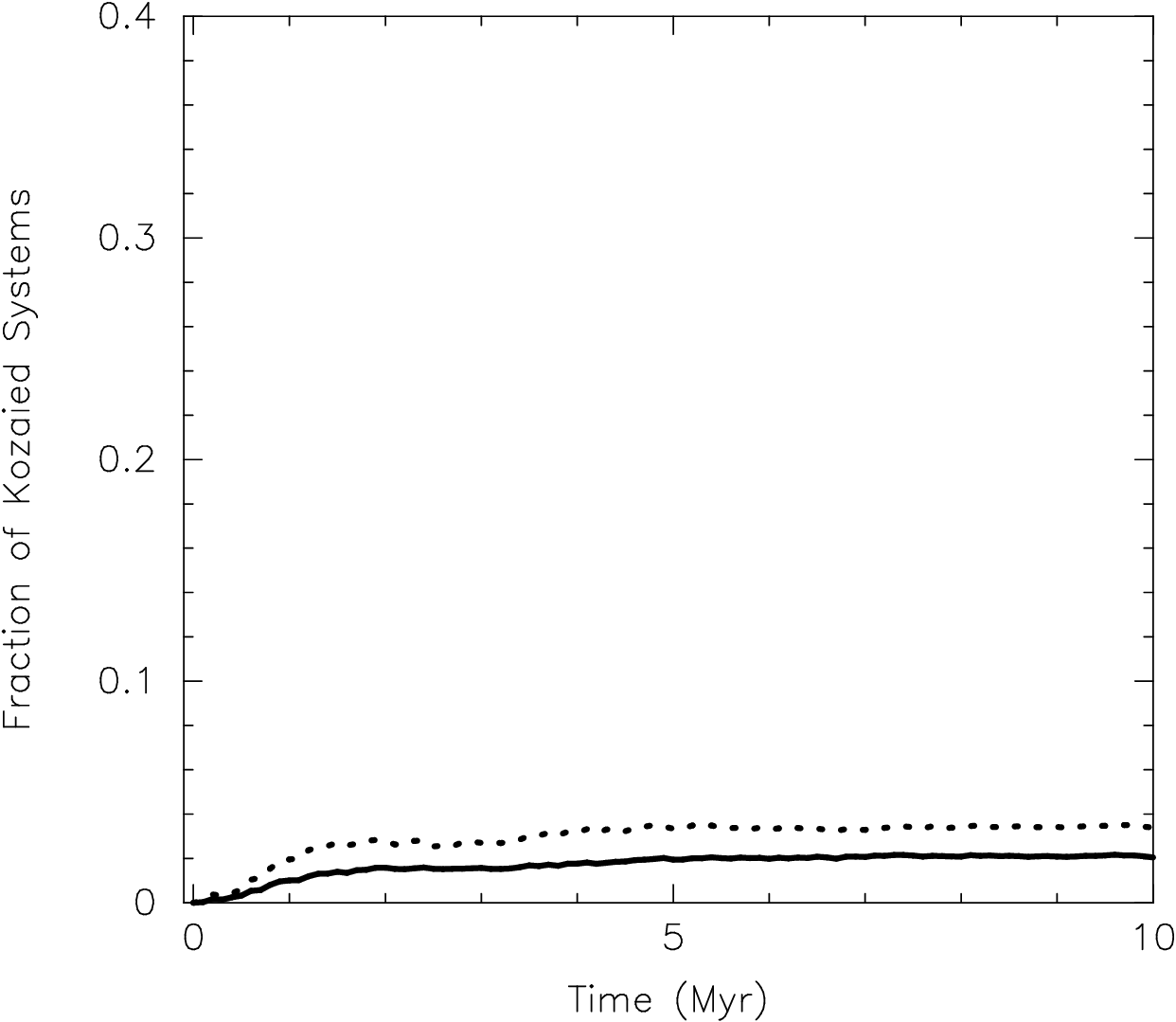}}}
\caption[bf]{The fraction of preserved binary systems where the mutual inclination between the two stars changes by more than 39$^\circ$ but less than 141$^\circ$, which are the required conditions for the von~Zeipel-Lidov-Kozai mechanism to operate. The solid line indicates the fraction of all binary systems that could be subjected to the vZLK mechanism, and the dotted line is the fraction of systems whose initial semimajor axis lies in the range that would be observed in most nearby star-forming regions (10 -- 1000\,au). Panel(a) shows the results for star-forming regions with a high degree of initial substructure ($D = 1.6$) and panel (b) shows the results for star-forming regions with no initial substructure ($D = 3.0$).} 
\label{kozai_frac}
  \end{center}
\end{figure*}

In Fig.~\ref{kozai_frac} we show the fraction of all preserved binary systems that could be subjected to the vZLK mechanism by the solid lines, and the fraction of systems with initial semimajor axes in the range 10 -- 1000\,au that could be   be subjected to the vZLK mechanism by the dotted lines. In the initially substructured star-forming regions, the high stellar density leads to 8\,per cent of all binaries, and 12\,per cent of 10 -- 1000\,au binaries being  subjected to the vZLK mechanism (Fig.~\ref{kozai_frac-a}), whereas the respective fractions are much lower (2\,per cent and nearly 4\,per cent) in the lower-density, smooth regions. These fractions are somewhat lower than those calculated in \citet{Plummer11} sphere clusters \citep{Parker09c}, but in that paper  binaries with separations greater than 100\,au only were considered (and these are more susceptible to dynamical encounters). Furthermore, \citet{Parker09c} also included binaries that formed during the simulation in their calculations, whereas we do not consider these systems in this paper.  

\subsection{Altered habitable zones}

\begin{figure}
  \begin{center}
%{\label{}\rotatebox{270}{\includegraphics[scale=0.35]{plot_Koz_frac_Or_C0p3F1p61pRmR10_10-1000au.ps}}}
\rotatebox{270}{\includegraphics[scale=0.35]{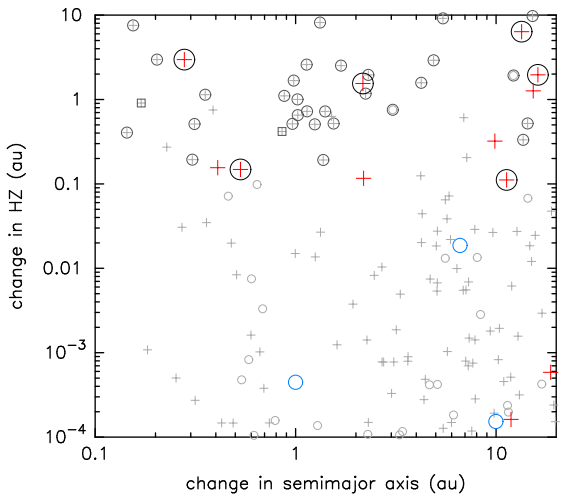}}
\caption[bf]{Change in habitable zone against the change in semimajor axis of binary stars in smooth ($D = 3.0$) star-forming regions. Red plus signs indicate binaries that have been hardened by a dynamical encounter, and blue circles indicate binaries that have been softened by a dynamical encounter. The points encompassed by a black circle are binaries for which the habitable zones overlap at periastron. The grey points are the data from \citet{Wootton19} for a similar initial binary population placed in a much more dense, substructured ($D = 1.6$) star-forming region.} 
\label{HZ_change}
  \end{center}
\end{figure}

In previous work \citep{Wootton19} we demonstrated that several binaries in each star-forming region could experience a hardening interaction that causes the two components to move closer together (especially at periastron). These simulations were of highly substructured ($D = 1.6$) regions with a population of field-like binaries. In Fig.~\ref{HZ_change} we show the results for a star-forming region with no substructure ($D = 3.0$). As these regions are less dense than their substructured counterparts, far fewer systems are altered such that previously distinct habitable zones around two component stars of a binary would merge. Systems where the habitable zones merge at periastron are indicated on Fig.~\ref{HZ_change} by an additional black circle. In this simulation, only systems that harden merge at periastron, and these systems are shown as the encircled plus symbols in Fig.~\ref{HZ_change}.

For comparison, we show via the faded symbols the results from the substructured regions first presented in \citet{Wootton19}. In the substructured regions, a total of 354 binary systems across a suite of 20 statistically identical simulations had their isophote habitable zones enlarged, whereas for the smooth regions it is only 75 systems (again, across 20 simulations). %We are at pains to stress that even if the isophote habitable zone increases in size due to an interaction, the binary will likely become more eccentric and therefore the area of the habitable zone that is predicted to host a stable planet could decrease. \citet{Georgakarakos19} repeated one of the calculations in \citet{Wootton19} to demonstrate that the dynamically stable part of the isophote habitable zone would actually decrease for the system in question, although this was also clearly pointed out in \citet{Wootton19}.
We stress that even if the isophote habitable zone increases in size due to an interaction, the binary will likely become more eccentric and therefore the area of the habitable zone that is predicted to host a stable planet could decrease. We note that \citet{Georgakarakos19} repeated one of the calculations in \citet{Wootton19}, confirming that the dynamically stable part of the isophote habitable zone would decrease for the system in question, in agreement with \citet{Wootton19}.

\subsection{Alternative binary populations}

Up to this point, we have assumed a field-like binary population as the initial binary population in our star-forming regions, in terms of the binary fraction and the initial semimajor axis distribution (both of which change as a function of the mass of the primary star, $m_p$). We also ran a set of simulations with the so-called `universal' initial binary population from \citet{Kroupa95a}, in which the initial binary fraction is unity and there is an excess of wide binary systems. There is considerable debate in the literature as to the validity of this model, with some authors arguing that then field binary population resembles the birth population \citep{King12a,King12b,Parker14d,Parker14e}, and other authors arguing that some dynamical processing must have occurred \citep{Marks14}.

Irrespective of that debate, it is important to recognise that binaries in the field may originate from star-forming regions with a multitude of different densities, which could have affected the binary populations in different ways.

We present the fractions of preserved binary systems from the \citet{Kroupa95a} `universal' population that could host stable planets at various distances on S-type orbits in Fig.~\ref{circum_prime_fracs_pres_universal} and the fractions of preserved binary systems that could host stable planets at various distances on P-type orbits in Fig.~\ref{circum_bin_fracs_pres_universal}. In contrast to the field population, the universal binary population leads to very similar semimajor axis distributions for binaries, irrespective of primary mass, and as such there are differences in the numbers of binaries that could host stable planets at the given distances when compared to the field populations.. 

As an example, the fraction of binaries that could host stable planets on S-type orbits at 1\,au is 50\,per cent for the Universal binary population, whereas it is 55\,per cent for the field-like populations. The trends are very similar for other planetary distances between the different binary populations for star-forming regions with the same initial properties. The differences between the populations are typically 5\,per cent, which is a similar to the difference between star-forming regions with different initial substructure but the same binary populations.

\begin{figure}
  \begin{center}
\rotatebox{270}{\includegraphics[scale=0.35]{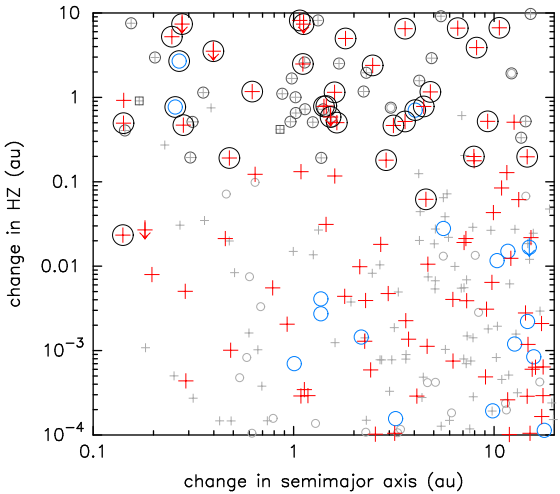}}
%{\includegraphics[scale=0.55]{plot_HZ_sims.ps}}
\caption[bf]{Change in habitable zone against the change in semimajor axis of binary stars in star-forming regions with the \citet{Kroupa95a} `universal' initial binary population. Red plus signs indicate binaries that have been hardened by a dynamical encounter, and blue circles indicate binaries that have been softened by a dynamical encounter. The points encompassed by a black circle are binaries for which the habitable zones overlap at periastron. The faded points are the data from \citet{Wootton19} for a similar initial binary population placed in a similar star-forming region but with a Field-like binary population.} 
\label{HZ_change_universal}
  \end{center}
\end{figure}

Similarly, the plots that show the change in the isophote habitable zones are qualitatively very similar. Similar total numbers of binaries have their habitable zones significantly altered (340 binaries in the universal population compared to 354 in the Field-like population, see Fig.~\ref{HZ_change_universal}) and the magnitude of these changes are also very similar (in part, because the binaries always have to be less than $\sim$10\,au apart for the habitable zones to change, and these close separation binaries have very similar properties in the field-like and universal populations). Systems where the habitable zones merge at periastron are indicated on Fig.~\ref{HZ_change_universal} by an additional black circle. In this simulation, the habitable zones have merged both in systems that have hardened (the encircled plus symbols) and systems that have softened (the circles within circles).

Secondly, much of our analysis concerns preserved binaries. By definition, these binaries are not broken apart by dynamical encounters (although they may have their orbital parameters altered) and such binaries tend to be in the semimajor axis range 0 -- $\sim$100\,au. In our two binary populations (Field-like and Universal), there is little difference in the properties of binaries with semimajor axes in this range, and so the main variable in our simulations remains the degree of stellar substructure.

\section{Discussion}
\label{discuss}

Our results show that changing the initial conditions of star-forming regions (i.e.\,\,the stellar density) leads to around a 5\,per cent difference in the fraction of binary systems that could host a stable planet. The results are fairly insensitive to the distance of the planet from the host star. For example, consider a planet on an S-type orbit at either 1\,au or 30\,au from the primary star. For a field-like binary population, the fraction of stable binary systems that could host a planet at 1\,au is 55\,per cent, and the fraction that could host a planet at 30\,au is around 4\,per cent for a dense substructured star-forming region. The corresponding stable fractions for lower density, smooth initial conditions are 60\,per cent and 8\,per cent.

Of the binary systems that are preserved (i.e.\,\,those where, if a planetary system were to form in the binary, the planets would still have two suns at the end of our simulations), there is very little change in the fraction of binaries that could host a stable planet over the duration of the simulation. Some binary systems do have their orbits altered such that planets could become unstable, but these tend to be compensated for by systems that go from being unstable to stable. 

A drastic change in the initial binary population also has a minimal effect on our results. Taking the dense, substructured star-forming regions as an example, when we adopt the binary population observed in the Galactic field in our initial conditions, the fraction of binaries that could host a stable planet at e.g.\,\,5\,au is constant at 20\,per cent. If we instead adopt the \citet{Kroupa95a} `universal' binary population, the fraction of binaries that could host a stable planet at 5\,au is 28\,per cent. 

When assessing the absolute numbers of possible stable planetary systems, we need to consider how many planets can form around both binary and single stars. \citet{Parker13c} assumed that a planet could form around every single star, and then calculated the total fraction of stable planetary systems for different assumed binary populations. More planets were stable in a field-like population because of the lower binary fraction (e.g.\,\,$f_{\rm bin} = 0.46$ for G-type stars) compared  to the \citet{Kroupa95a} population, which assumes $f_{\rm bin} = 1.00$ for all stars (for planets at 5\,au on S-type orbits, the stable fractions in \citet{Parker13c} were 75\,per cent for the field-like population, and 60\,per cent for the universal population). Therefore, even with the rather extreme dynamical environments we model is this paper, the dominant factor for whether planets can be stable in a binary population is the characteristics of that population (binary fraction, and to a lesser extent the distributions of the various orbital parameters). 

 If we assume that \emph{all} pre-main sequence single stars have a disc from which planets can form, then there are several processes that can affect the long-term stability of the planetary system. 
 
 External photoevaporation from massive stars \citep[e.g.][]{Scally01,Adams04,Nicholson19a} can truncate or destroy the gaseous component of a protoplanetary disc \citep{Haworth18b}, but doesn't necessarily preclude the formation of terrestrial planets. Similarly, truncation of discs via direct encounters with passing stars can also be detrimental \citep[e.g.][]{Vincke16,Zwart16}, but these effects are limited to only the most dense star-forming regions (and not all stars within these regions).
 
Once any planets have formed, dynamical encounters with passing stars can create free-floating planets \citep{Hurley02,Parker12a} and lead to the disruption of orbital parameters \citep[e.g.][]{Bonnell01,Adams06}. However, these processes affect at most around 50\,per cent of all single stars, whereas our calculations indicate that the fraction of \emph{unstable} binaries that could not host S-type planets at distances of 5\,au or more is greater than 80\,per cent (depending on the location of the planet). 

Therefore, it appears that the dominant factor for assessing the stability of planetary systems is the overall fraction of single stars at birth (assuming that the processes of disc destruction/truncation, and planetary system modification are not more prevalent for planets around single stars).

\section{Conclusions}
\label{conclude}

We present $N$-body simulations of the dynamical evolution of star-forming regions that contain populations of primordial binary systems to determine whether the fraction of binaries that could host planetary systems is affected by dynamical encounters. Our conclusions are the following:

(i) The fraction of binaries that could host a stable planet at a given distance is relatively constant for the duration of the evolution of the star-forming region. Systems that could host a stable planet within 10\,au of the primary star before dynamical evolution are unlikely to be broken apart by a strong dynamical encounter. 

(ii) Similarly, binary systems that are preserved throughout the duration of the simulation can have their orbital parameters altered, but rarely in such a way that a previously stable planet would be made unstable. These changes to the orbital properties of some binaries can affect the long-term stability of systems by e.g. inducing von~Zeipel-Lidov-Kozai cycles, or increase the size of the habitable zones in binary systems, but these processes tend to be confined to only the most dense star-forming regions. 

(iii) The initial binary population is a more dominant factor than the initial density of the star-forming region. For example, if the binary fraction is low, we would expect more stable planets as there would be fewer binary systems, whereas a high binary fraction would preclude the formation of stable planets on certain orbits (depending on the properties of the binary system in question).\\

Taken together, we suggest that the stability of planets on both S-type or P-type orbits in and around binary star systems is almost independent of any dynamical evolution that may occur in their birth star-forming region. The main factor is the initial properties of the binary population. If the birth population were universal, i.e. the same in each star-forming region \citep[as postulated by][]{Kroupa95a}, then one can in principle predict the absolute fraction of stable planetary systems that could form in and around binary systems. 

\section*{Acknowledgements}

We thank the anonymous referee and scientific editor for helpful feedback that significantly improved this paper. RJP acknowledges support from the Royal Society in the form of a Dorothy Hodgkin Fellowship. We thank Helena Gibbon for creating a video abstract for this paper (available at: \href{https://youtu.be/76kANnTK9-s}{https://youtu.be/76kANnTK9--s}).

\section*{Data availability statement}

The data used to produce the plots in this paper will be shared on reasonable request to the corresponding author.

\bibliographystyle{mnras}  
\bibliography{general_ref}

\appendix

\section{Stability fractions for preserved binaries}
\label{appendix:preserved}

 We now repeat the analysis of the stability fraction of planets at 1, 5, 10, 30, 50 and 100\,au in and around binaries but limit ourselves to binary systems that are preserved throughout the duration of the simulation. These are binaries that are present in the initial conditions, and every subsequent snapshot and who retain their birth partners throughout. 

\begin{figure*}
  \begin{center}
\setlength{\subfigcapskip}{10pt}
%\hspace*{0.3cm}\subfigure[$D = 1.6$]{\label{}\rotatebox{270}{\includegraphics[scale=0.35]{plot_HW_S_pres_Or_C0p3F1p61pB_F10.ps}}}
%\hspace*{0.5cm}\subfigure[$D = 3.0$]{\label{}\rotatebox{270}{\includegraphics[scale=0.35]{plot_HW_S_pres_Or_C0p3F3p01pB_F10.ps}}}
\hspace*{0.3cm}\subfigure[$D = 1.6$]{\label{circum_prime_fracs_pres-a}\rotatebox{270}{\includegraphics[scale=0.35]{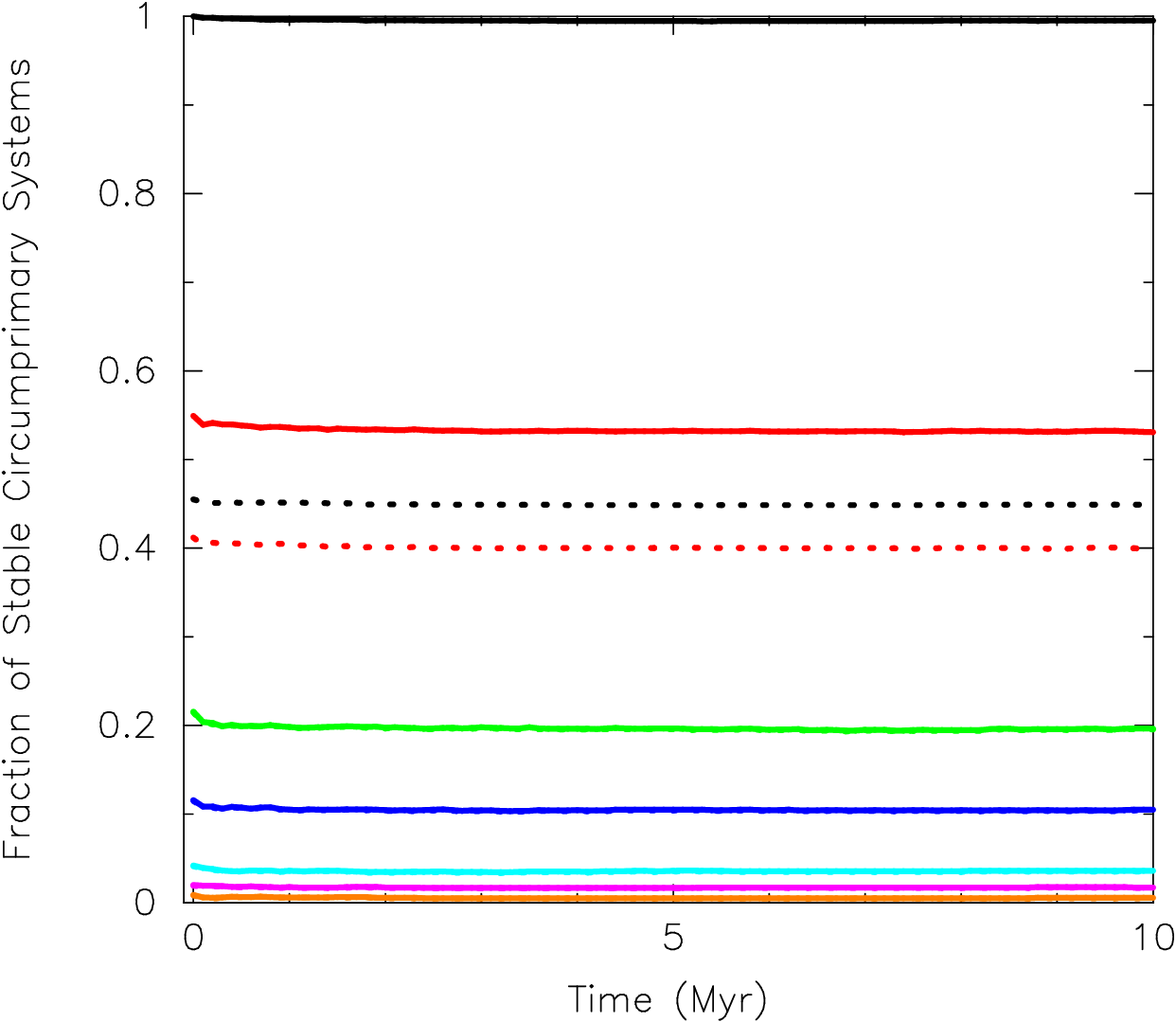}}}
\hspace*{0.5cm}\subfigure[$D = 3.0$]{\label{circum_prime_fracs_pres-b}\rotatebox{270}{\includegraphics[scale=0.35]{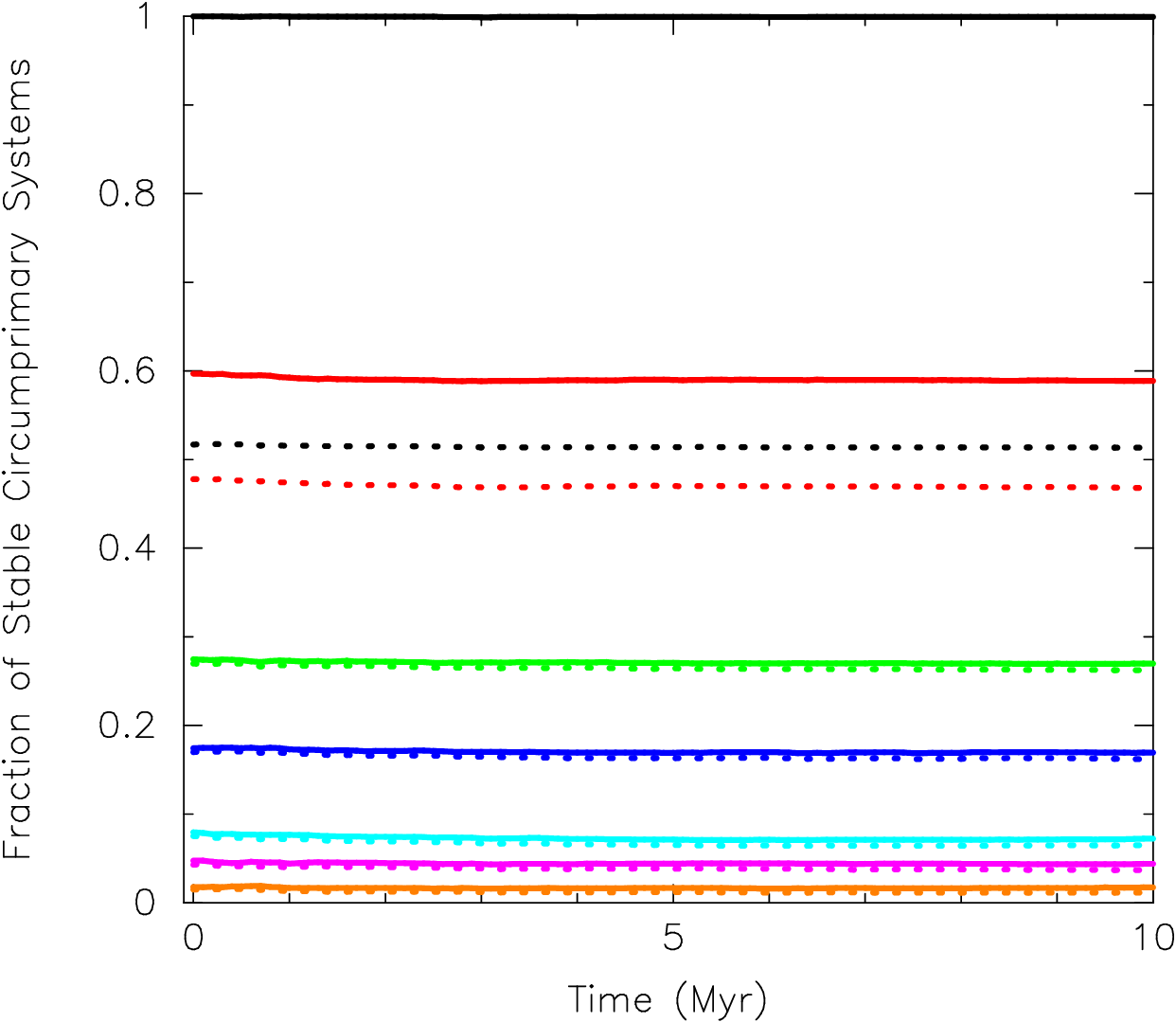}}}
\caption[bf]{Fraction of \emph{preserved} binary systems in which planets on S-type (circumprimary) orbits are stable according to the \citet{Holman99} criteria. The solid lines show the fraction of the preserved binary systems (primordial binaries that are present at all times during the simulation) that could host a stable planet, whereas the dotted lines show the corresponding fractions of preserved binary systems that have initial semimajor axes in the range 10--1000\,au that could host stable circumprimary planets. The black lines indicate the fraction of binary systems that could host a planet within some distance from the primary star (and as such is nearly unity when we do not apply a cut to the binary semimajor axes); the red lines indicate the fraction of binaries that could host a stable planet at 1\,au from the primary star; the green lines indicate the fraction of binaries that could host a planet at 5\,au; the dark blue lines indicate the fraction of binaries that could host a planet at 10\,au; the cyan lines indicate the fraction of binaries that could host a planet at 30\,au; the magenta lines indicate the fraction of binaries that could host a planet at 50\,au and the orange lines indicate the fraction of binaries that could host a planet at 100\,au from the primary star. In panel (a) we show the results for a evolution of a population of field-like binaries in a highly substructured (fractal dimension $D = 1.6$) star-forming region, and in panel (b) we show the results for a star-forming region with no substructure, i.e.\,\,smooth (fractal dimension $D = 3.0$).} 
\label{circum_prime_fracs_pres}
  \end{center}
\end{figure*}

\begin{figure*}
  \begin{center}
\setlength{\subfigcapskip}{10pt}
%\hspace*{0.3cm}\subfigure[$D = 1.6$]{\label{}\rotatebox{270}{\includegraphics[scale=0.35]{plot_HW_P_pres_Or_C0p3F1p61pB_F10.ps}}}
%\hspace*{0.5cm}\subfigure[$D = 3.0$]{\label{}\rotatebox{270}{\includegraphics[scale=0.35]{plot_HW_P_pres_Or_C0p3F3p01pB_F10.ps}}}
\hspace*{0.3cm}\subfigure[$D = 1.6$]{\label{circum_bin_fracs_pres-a}\rotatebox{270}{\includegraphics[scale=0.35]{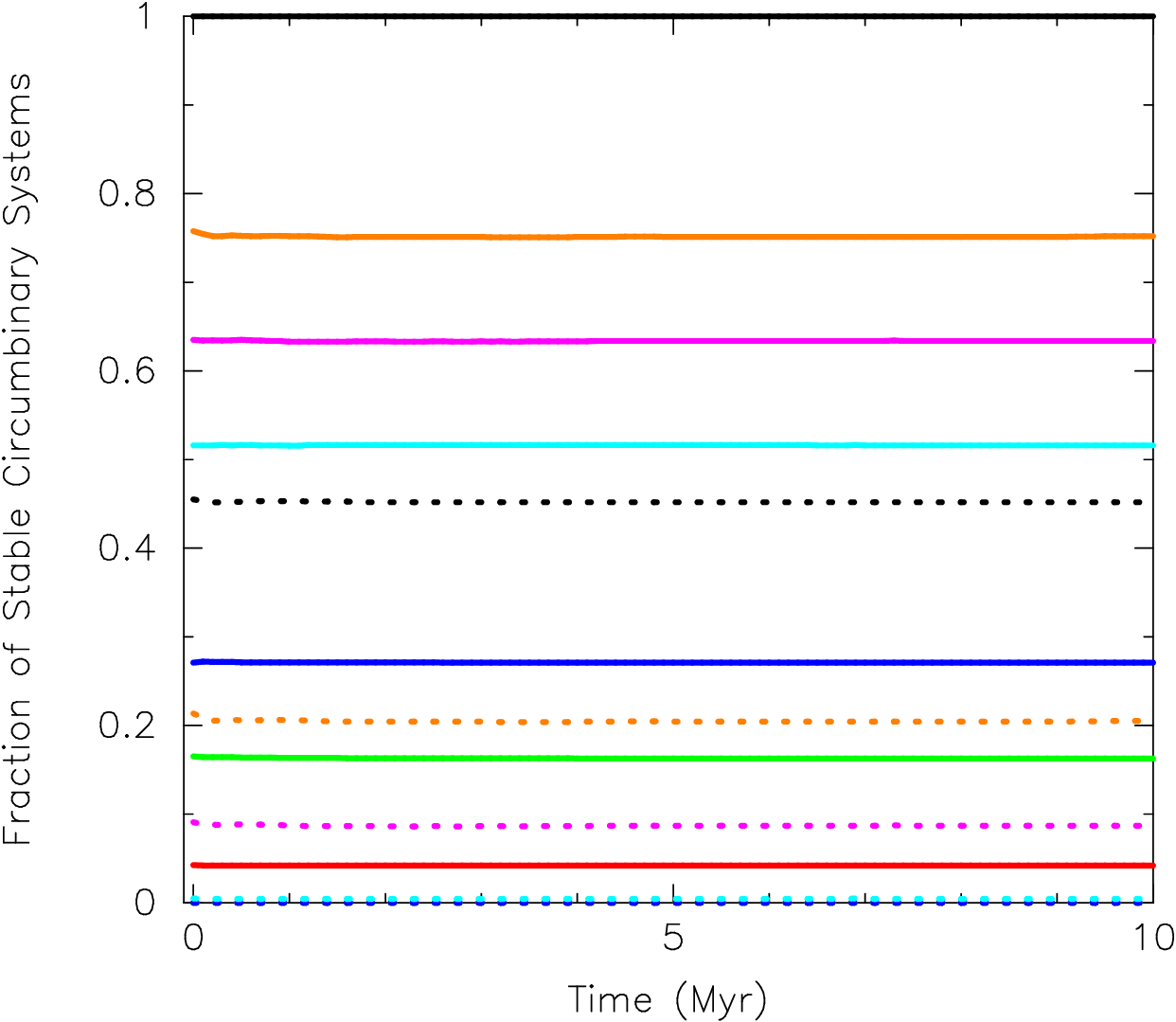}}}
\hspace*{0.5cm}\subfigure[$D = 3.0$]{\label{circum_bin_fracs_pres-b}\rotatebox{270}{\includegraphics[scale=0.35]{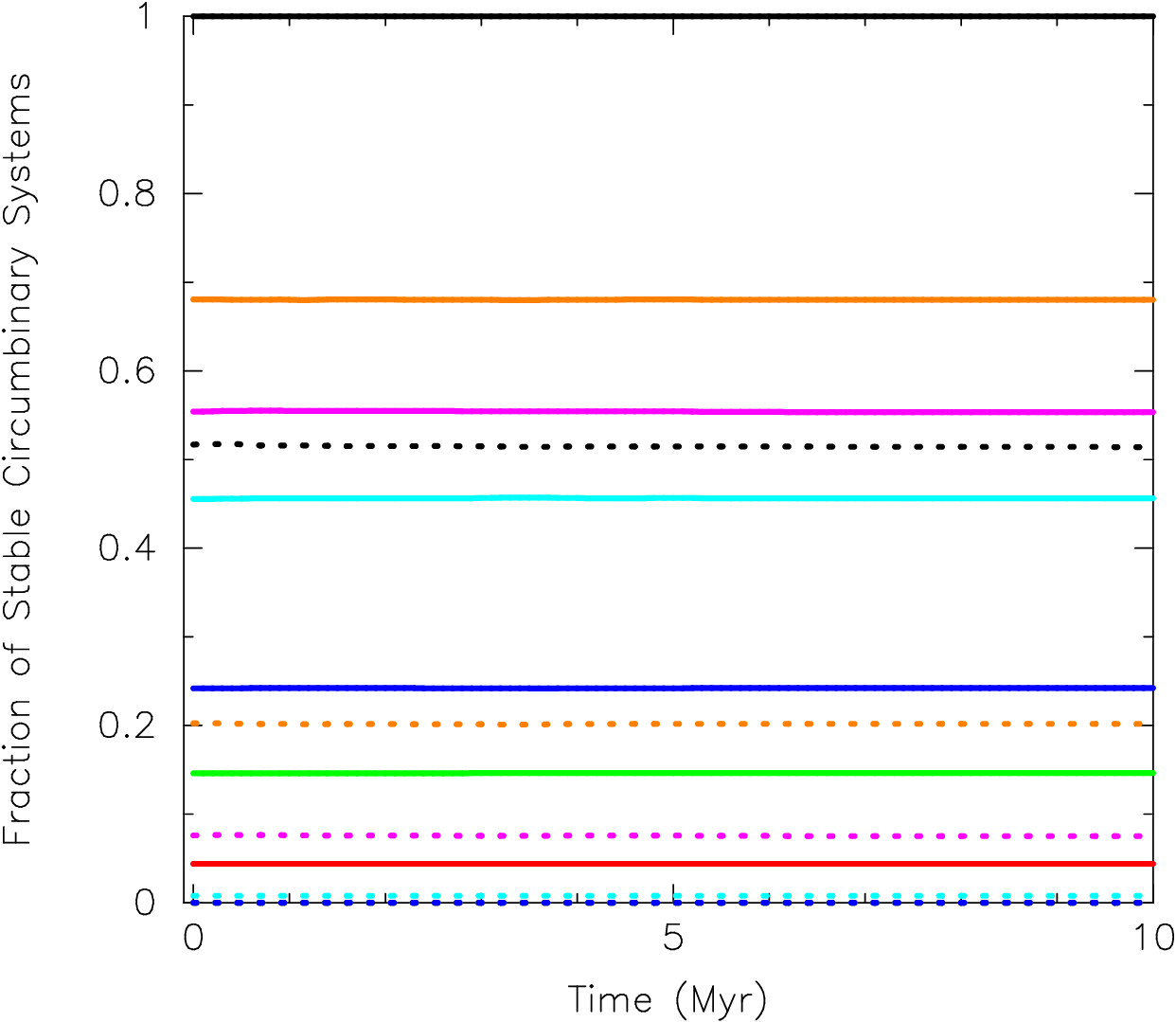}}}
\caption[bf]{Fraction of \emph{preserved} binary systems in which planets on P-type (circumbinary) orbits are stable according to the \citet{Holman99} criteria.  The solid lines show the fraction of the preserved binary systems (primordial binaries that are present at all times during the simulation) that could host a stable planet, whereas the dotted lines show the corresponding fractions of preserved binary systems that have initial semimajor axes in the range 10--1000\,au that could host stable circumbinary planets. The black lines indicate the fraction of binary systems that could host a planet within some distance from the binary (and as such is nearly unity when we  do not apply a cut to the binary semimajor axes); the red lines indicate the fraction of binaries that could host a stable planet at 1\,au from the binary; the green lines indicate the fraction of binaries that could host a planet at 5\,au; the dark blue lines indicate the fraction of binaries that could host a planet at 10\,au; the cyan lines indicate the fraction of binaries that could host a planet at 30\,au; the magenta lines indicate the fraction of binaries that could host a planet at 50\,au and the orange lines indicate the fraction of binaries that could host a planet at 100\,au from the binary system. In panel (a) we show the results for a evolution of a population of field-like binaries in a highly substructured (fractal dimension $D = 1.6$) star-forming region, and in panel (b) we show the results for a star-forming region with no substructure, i.e.\,\,smooth (fractal dimension $D = 3.0$).} 
\label{circum_bin_fracs_pres}
  \end{center}
\end{figure*}

We show the evolution of the fraction of stable preserved binaries that could host a planet at various distances from the primary star in Fig.~\ref{circum_prime_fracs_pres} and the evolution of the fraction of stable preserved binaries that could host a planet at various distances from the binary in Fig.~\ref{circum_bin_fracs_pres}.

\section{Stability fractions for alternative binary populations}

\begin{figure}
  \begin{center}
%{\label{}\rotatebox{270}{\includegraphics[scale=0.35]{plot_Koz_frac_Or_C0p3F1p61pRmR10_10-1000au.ps}}}
\rotatebox{270}{\includegraphics[scale=0.35]{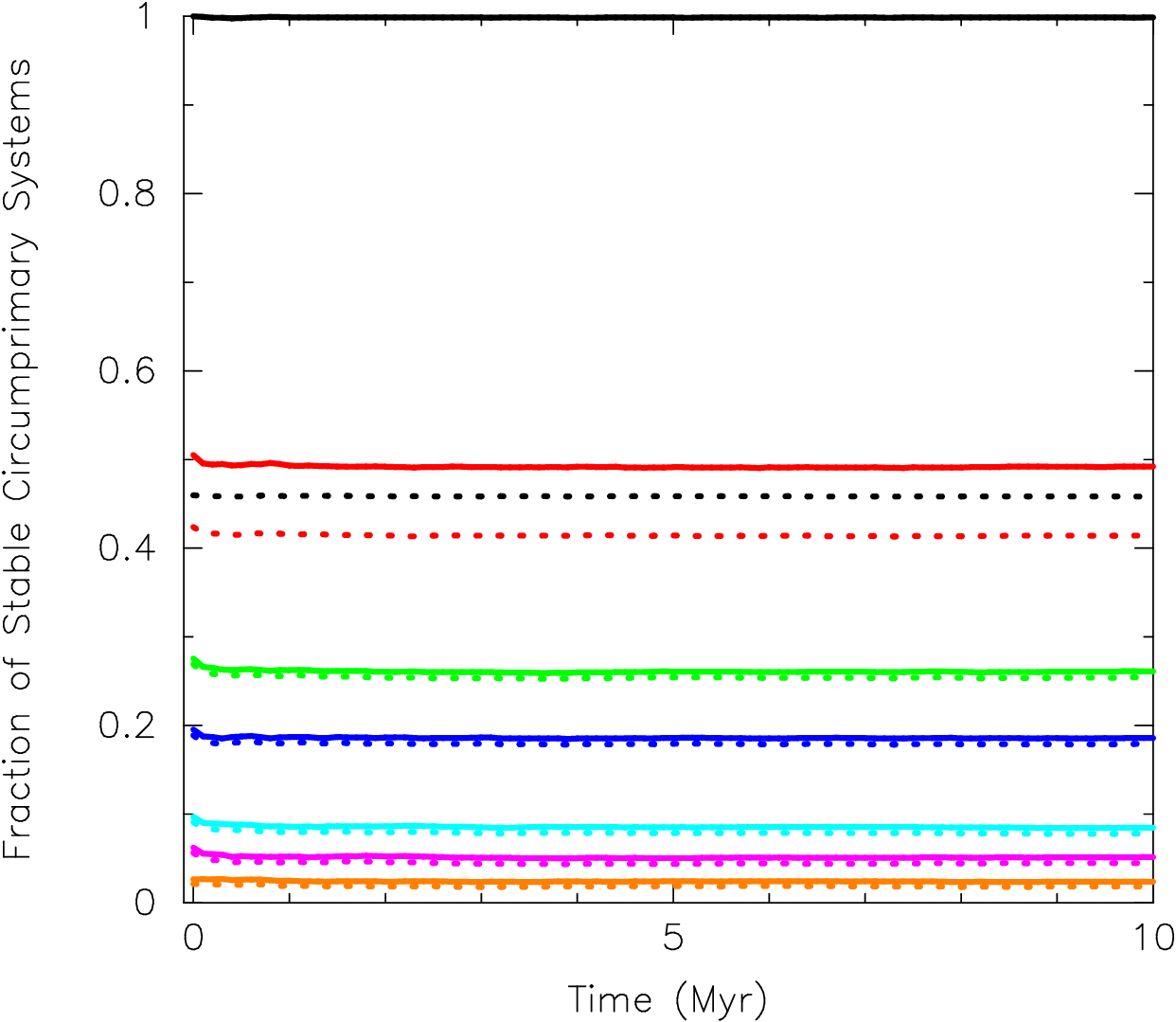}}
\caption[bf]{Fraction of \emph{preserved} binary systems in which planets on S-type (circumprimary) orbits are stable according to the \citet{Holman99} criteria for the \citet{Kroupa95a} `Universal' binary population (this plot should be directly compared to Fig.~\ref{circum_prime_fracs_pres-a}). The solid lines show the fraction of the preserved binary systems (primordial binaries that are present at all times during the simulation) that could host a stable planet, whereas the dotted lines show the corresponding fractions of preserved binary systems that have initial semimajor axes in the range 10--1000\,au that could host stable circumprimary planets. The black lines are the fraction of binaries that could host a planet at some distance, whereas the red, green, dark blue, cyan, magenta and orange lines are the fractions of binaries that could host stable planets at 1, 5, 10, 30, 50 and 100\,au, respectively. }
%The black lines indicate the fraction of binary systems that could host a planet within some distance from the primary star (and as such is nearly unity when we don't apply a cut to the binary semimajor axes); the red lines indicate the fraction of binaries that could host a stable planet at 1\,au from the primary star; the green lines indicate the fraction of binaries that could host a planet at 5\,au; the dark blue lines indicate the fraction of binaries that could host a planet at 10\,au; the cyan lines indicate the fraction of binaries that could host a planet at 30\,au; the magenta lines indicate the fraction of binaries that could host a planet at 50\,au and the orange lines indicate the fraction of binaries that could host a planet at 100\,au from the primary star. .} 
\label{circum_prime_fracs_pres_universal}
  \end{center}
\end{figure}

\begin{figure}
  \begin{center}
%{\label{}\rotatebox{270}{\includegraphics[scale=0.35]{plot_Koz_frac_Or_C0p3F1p61pRmR10_10-1000au.ps}}}
\rotatebox{270}{\includegraphics[scale=0.35]{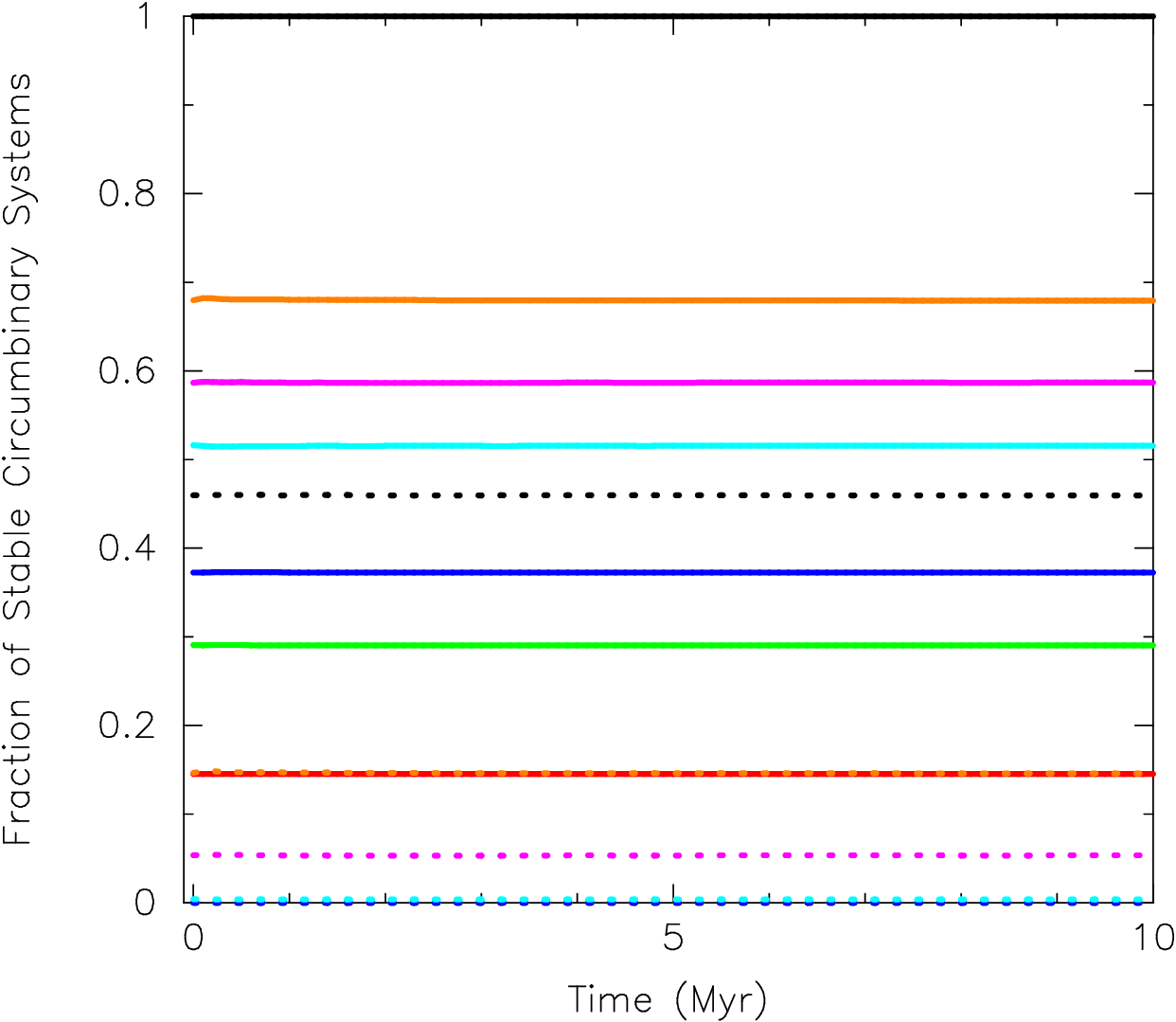}}
\caption[bf]{Fraction of \emph{preserved} binary systems in which planets on P-type (circumbinary) orbits are stable according to the \citet{Holman99} criteria for the \citet{Kroupa95a} `Universal' binary population (this plot should be directly compared to Fig.~\ref{circum_bin_fracs_pres-a}).  The solid lines show the fraction of the preserved binary systems (primordial binaries that are present at all times during the simulation) that could host a stable planet, whereas the dotted lines show the corresponding fractions of preserved binary systems that have initial semimajor axes in the range 10--1000\,au that could host stable circumbinary planets. The black lines are the fraction of binaries that could host a planet at some distance, whereas the red, green, dark blue, cyan, magenta and orange lines are the fractions of binaries that could host stable planets at 1, 5, 10, 30, 50 and 100\,au, respectively.} 
%The black lines indicate the fraction of binary systems that could host a planet within some distance from the binary (and as such is nearly unity when we don't apply a cut to the binary semimajor axes); the red lines indicate the fraction of binaries that could host a stable planet at 1\,au from the binary; the green lines indicate the fraction of binaries that could host a planet at 5\,au; the dark blue lines indicate the fraction of binaries that could host a planet at 10\,au; the cyan lines indicate the fraction of binaries that could host a planet at 30\,au; the magenta lines indicate the fraction of binaries that could host a planet at 50\,au and the orange lines indicate the fraction of binaries that could host a planet at 100\,au from the binary system.} 
\label{circum_bin_fracs_pres_universal}
  \end{center}
\end{figure}

In Figs.~\ref{circum_prime_fracs_pres_universal}~and~\ref{circum_bin_fracs_pres_universal} we show the evolution of the stability fractions for planets at 1, 5, 10, 30, 50 and 100\,au for binaries drawn from the \citet{Kroupa95a} `Universal' binary population. 

We show the evolution of the fraction of stable preserved binaries that could host a planet at various distances from the primary star in Fig.~\ref{circum_prime_fracs_pres_universal} and the evolution of the fraction of stable preserved binaries that could host a planet at various distances from the binary in Fig.~\ref{circum_bin_fracs_pres_universal}.

\label{lastpage}

\end{document}